\documentstyle[sprocl]{article}

\def\simge{\mathrel{%
   \rlap{\raise 0.511ex \hbox{$>$}}{\lower 0.511ex \hbox{$\sim$}}}}
\def\simle{\mathrel{
   \rlap{\raise 0.511ex \hbox{$<$}}{\lower 0.511ex \hbox{$\sim$}}}}

\def\Journal#1#2#3#4{{#1} {\bf #2}, #3 (#4)}

\def\be{\begin{equation}}
\def\ee{\end{equation}}
\def\bea{\begin{eqnarray}}
\def\eea{\end{eqnarray}}
\def\ba{\begin{array}} 
\def\ea{\end{array}}   

\begin{document}

\title{\large \bf THE ~STANDARD ~MODEL ~AND ~BEYOND}

\author{\vskip .2truecm P. FAYET}

\address{
Laboratoire de Physique Th\'eorique de l'Ecole Normale 
Sup\'erieure\,\footnote{ Unit\'e Propre du CNRS, associ\'ee \`a 
l'Ecole Normale Sup\'erieure et \`a l'Universit\'e Paris-Sud.\\ 
\vskip -.1truecm
\small{\bf{LPTENS-98/45 \, hep-ph/9812300}}  \ \,
{\small {Based in particular on a talk at the Conf. 
``Parity Violation in Electron-Hadron Electroweak Interactions'' 
(oct. 1997, Paris).}}
}, 
\\ 
24 rue Lhomond,
75231 Paris Cedex 05, France,
\\E-mail: fayet@physique.ens.fr}

\maketitle\abstracts{
\vskip .7truecm
We start with a brief presentation of the Standard Model 
and the weak neutral current, in view of a discussion 
of parity-violation in electron-hadron electroweak interactions.
We then discuss some limitations of this model, motivating 
the consideration of larger frameworks  such as grand-unification and 
supersymmetry. We also comment about the possible effects 
of additional {\it \,light\,} neutral gauge bosons.}

{\section{The Standard Model and the Weak Neutral Current}}
\label{sec:sm}

\vskip .1truecm

What is the Standard Model of Particles and Interactions\,?  
And, at first, what are the particles and interactions we are talking about\,?
Does this model give a satisfactory description of the real world\,?
What did we learn from the discovery and study of Parity Violation in 
electron-hadron interactions, the subject of this Conference\,?
Can we be satisfied with this Standard Model, or do we have 
reasons to go beyond it\,? 
If so, what sort of ``new physics'' could we expect, 
and how could it be tested\,?  Here are some of the questions 
that we would like to address \,-- although rather briefly --\, 
in this general introductory talk.

\vskip.2truecm

It is well \,known that the Standard Model provides a very good description 
of particle interactions. Here we are talking about weak, 
electromagnetic and strong interactions of known particles, 
essentially leptons and hadrons, the latter being 
strongly-interacting particles built from quarks and antiquarks.
There are three charged leptons, the electron, the muon and the tau lepton, 
which all have their associated neutrinos, $\,\nu_e\,$,
$\,\nu_\mu\,$ and $\,\nu_\tau$. \,We also know six sorts of quarks, 
after the top quark was finally discovered at Fermilab in 1994. 
They have electrical charges $\,\frac{2}{3}\,$ and $\,-\,\frac{1}{3}\,$, 
~and may be associated two by two, as indicated in Table \ref{tab:leptonq}.
\begin{table}[t]
\caption{The three families of leptons and quarks.
Left-handed lepton and quark fields transform as members of $\,SU(2)\,$ 
electroweak doublets, while their right-handed counterparts are singlets.
\label{tab:leptonq}
}
\vspace{0.1cm}
\begin{center}
\begin{tabular}{|ccccc|}
\hline  &&&& \\
\ \ 6\ \,leptons: \ &
$\left( \ba{c} \nu_e \\ e^-  \ea \right)    $ & 
$\left( \ba{c} \nu_\mu \\ \mu^- \ea\right)  $ &
$\left( \ba{c} \nu_\tau \\ \tau^- \ea\right)$ & \\  
&&&& \\
\ \ 6\ \,quarks: \ &
$\left( \ba{c} \,u\, \\ \,d \,\ea \right) $ &
$\left( \ba{c} \,c\, \\ \,s\, \ea \right) $ &
$\left( \ba{c} \,t\, \\ \,b\, \ea \right) $ &  \\   &&&& \\ \hline    
\end{tabular}
\end{center}
\end{table}
We leave momentarily aside the fourth interaction, gravitation, 
very well described \,-- although only at the classical level --\,
by Einstein's theory of General Relativity.
At the level of individual particles it is extremely weak 
compared to the three other interactions, at all energies 
in which we shall be interested, as indicated by the extremely small 
value of the Newton constant, 
$\ G_{\footnotesize\hbox{Newton}}\ \simeq \ 10^{-38}\ \ \hbox{GeV}^{-2}\,$.

\vskip.2truecm

Weak, electromagnetic and strong interactions are now well understood 
from the exchanges of spin-1 mediators known as 
gauge bosons, between the ``constituents of matter'', 
as are often called the spin-$\frac{1}{2}\,$ leptons and quarks.
The spin-1 mediators are the famous gauge bosons of the Standard Model: 
\hbox{the {\it \,photon}}, \,whose exchanges between charged particles are 
responsible for the electromagnetic interactions. 
The eight {\it \,gluons\,} of Quantum Chromodynamics (QCD), 
whose exchanges between ``colored'' quarks 
(each quark appearing as a triplet of a color $\,SU(3)\,$ gauge group) 
are responsible for the strong interactions, binding  together quarks 
three by three to form protons \,($\,uud\,$)\, and neutrons \,($\,ddu\,$).
And the {\it \,intermediate gauge bosons} $\ W^\pm\,$ and $\,Z$, ~very heavy
 \,-- approximately \,80\, and \,91\, GeV$/c^2$, ~respectively --\,
responsible for the various ``charged-current'' and ``neutral-current'' 
weak-interaction processes\,\footnote{ More precisely, currently used 
values of the $\,W^\pm$ and $\,Z\,$ masses 
are \hbox{$\,m_W = 80.4 \,\pm \,.1$} $\hbox{GeV}/c^2$, ~and 
$\ m_Z = 91.187\, \pm .003\ \hbox{GeV}/c^2$.
~The fine-structure constant of electromagnetic interactions is 
$\,\alpha = e^2/4 \pi \epsilon_0 \hbar c \simeq 1/137.036$; 
its value at the $\,Z\,$ mass is given by 
$\,\alpha^{-\,1}(m_Z) = 128.9 \pm .1\,$.
~The Fermi coupling constant of the weak interactions is
$\,G_F\,= \, (1.166\,39 \,\pm \,.000 \, 01)\ \,10^{-5}\ \,
\hbox{GeV}^{-2}\ (\hbar c)^3$;
and the electroweak mixing angle, to be introduced later
(defined by $\ \tan \theta =\,g'/g\ $ with $\,g\,$ and $\,g'\,$ 
evaluated at the $\,Z\,$ mass) is given
by $\,\sin^2 \theta\,\simeq\,.231 \,4 \pm .000 \,3$. 
\smallskip
\par 
The QCD strong interaction coupling $\,\alpha_s = g_s^{\,2}/4\pi\,$ 
``runs'' with the energy, decreasing from large values at low energies, 
down to $\,\alpha_s(m_Z)\,\simeq\,.12$.
~The Newton constant of gravitational interactions is
$\ G_N\,\simeq \,(.670\,7 \pm\, .000 \,1)\ \,10^{-38}\ \ 
(\hbox{GeV}/c^2)^{-2}\ (\hbar c)\,$,
which corresponds to a ``Planck mass'', 
$\ m_P\,=\,\sqrt{\,\hbar c/G_N}\ 
\simeq\, 1.221\ \,10^{19}\ \hbox{GeV}/c^2\,\simeq \, 2.177\ \,10^{-5}$ g,
~and a ``Planck length'' 
$\ L_P\,\simeq\,\hbar/\, (m_P \,c)\,
\simeq \, 1.616 \ \,10^{-33}$ cm.
}.
\linebreak
These $\,W^\pm\,$ and $\,Z\,$ appear, together with the photon, 
as the four gauge bosons of the $\,SU(2) \times U(1)\,$ gauge symmetry group 
of the electroweak interactions~\cite{sm1,sm2}.
This theory was constructed in the sixties and early seventies, 
by extending to weak and strong interactions the ideas of gauge invariance
underlying the very successful theory of quantum electrodynamics (QED).
This one is the prototype of a renormalisable theory, in which quantum effects
can be computed in a consistent way and are found to be finite, 
order by order in perturbation theory.

\vskip.2truecm

If we go back to the sixties, a severe problem was posed by 
the definition of a quantum theory for the weak interactions. 
While these interactions were well described, at the classical level, 
by the Fermi theory involving local products of four fermion fields 
(times a coupling constant 
proportional to the Fermi constant, 
$\ G_F \simeq \,10^{-5}\ \,\hbox{GeV}^{-2}$),
~it was not possible to compute higher-order quantum corrections 
to weak-interaction processes, found to be severely divergent. 
The solution to this problem went through the understanding 
of (charged-current) weak interactions as due to the exchanges of charged heavy
intermediate bosons named $\ W^+$ and $\,W^-$, 
~coupled, proportionally to some dimensionless constant $\,g$, 
~to a linear combination of charged vector and axial-vector currents.
The Fermi constant $\,G_F\,$ may then be expressed
in terms of this coupling $\,g\,$ and the mass of the intermediate bosons 
$\,W^\pm$, ~according to the formula 
\be
\frac{G_F}{\sqrt 2} \ \ =\ \ \frac{g^2}{8\ m_W^{\ 2}}\ \ .
\ee

\vskip.2truecm

But this was not enough to provide a sensible quantum theory.
It appeared also necessary to view these spin-1 intermediate bosons 
$\,W^+\,$ and $\,W^-$ 
\linebreak
as gauge bosons coupled to charged currents, 
with a spontaneous symmetry breaking mechanism 
at work to provide these gauge bosons a mass
(otherwise they would remain exactly massless, just like the photon, 
as a consequence of the conserved gauge symmetry 
of the theory)~\cite{sm1}$^-$\,\cite{higgs}.
The corresponding gauge group had to be related with the 
$\,U(1)\,$ gauge group of quantum electrodynamics, 
since the $\,W^\pm$'s are charged and must interact with the photon.
Using simply $\,SU(2)\,$ (with $\,U(1)_{\hbox{\tiny{QED}}}\,$ as a subgroup) 
would not have provided a really satisfactory solution; 
nowadays we can simply see, for example, that it would not allow for quarks
having fractional electrical charges $\,\frac{2}{3}$ and $\,-\,\frac{1}{3}\,$.
~This led to the consideration by Glashow, Salam, Ward and 
Weinberg of $\ SU(2) \times U(1)\,$ as the proposed 
(spontaneously broken) gauge symmetry group
of electromagnetic and weak interactions. These interactions become,
in some sense, partly unified within the framework of a single 
``electroweak'' theory.

\vskip.2truecm

But $\,SU(2) \times U(1)\,$ has four generators,
the ``weak-isospin'' generators $\,T_1,\ T_2\,$ and $\,T_3$, 
~and the ``weak hypercharge'' $\,Y$, ~not only three.
This required, if the idea was to be right, the existence of an 
additional neutral gauge boson, the $\,Z\,$ \,-- which had to be very heavy --\,
coupled to a new 
``weak neutral current''. At the same time the massless photon field $\,A\,$ 
is coupled to the electromagnetic current, associated with
the electric charge now expressed, in terms of the elementary unit $\,e$, 
~by 
\be
\label{Q}
Q\ \ =\ \ T_3\ +\ \frac{Y}{2}\ \ .
\ee

\vskip.2truecm

More precisely, if we denote by $\,W_1,\ W_2, \ W_3\ $ and $\,B\ $ 
the four gauge fields of $\,SU(2) \times U(1)\,$,
~the charged $\,W\,$ fields are given by 
\be
W^\pm\ \ =\ \ \frac{W_1\,\mp\,i\,W_2}{\sqrt 2}\ \ ,
\ee
while the fields of the new weak neutral boson $\,Z\,$  
and of the photon $\,\gamma\,$ are given by the two orthogonal linear 
combinations
\be
\left\{\ \
\ba{ccccc}
Z \ \ &=& \ \ \cos \theta \ \ W_3&\ -\ & \sin \theta \ \ B \ \ , \vspace{2mm} \\
A \ \ &=& \ \ \sin \theta \ \ W_3&\ +\ & \cos \theta \ \ B \ \ .\\  
\ea \right.
\ee
$\theta\,$ is the electroweak mixing angle, fixed in terms of the 
two $\,SU(2)\,$ and $\,U(1)$ gauge coupling constants $\,g\,$ and $\,g'\,$ 
by 
\be
\tan \,\theta\ \ =\ \ \frac{g'}{g}\ \ ,
\ee
the elementary charge being given by 
\be
\label{e}
e\ =\ g\ \sin\,\theta\ =\ g'\ \cos\,\theta\ \ .
\ee

\vskip.2truecm

The new weak neutral current coupled to the $\,Z\,$ boson is then 
expressed\,\footnote{ The couplings of the neutral gauge bosons 
to chiral quark and lepton fields 
$ \ \psi_{L,R}\,=\,\frac{1\,\mp\,\gamma_5}{2}\,\psi$ 
are obtained from the Lagrangian density
\bea
{\cal L}\ \ =\ \ \overline{\psi_{L,R}} \ \ 
[\ i\,\partial_\mu\,-\,(\,g\ T_3\ W_{\mu \,3}\ +\ \frac{g'}{2}\ Y\ B_\mu\,)\ ]
\ \,\gamma^\mu\ \psi_{L,R} \ \ ,
\nonumber
\eea
by rediagonalizing the couplings of the neutral gauge fields 
$\,W_{\mu \,3}\,$ and $\,B_\mu\,$ 
to the corresponding weak isospin and weak hypercharge currents, as follows:
\bea
g\ \,J^\mu_{\ \,3}\ W_{\mu \,3}\ +\ \frac{g'}{2}\ \,J^\mu_{\ \,Y}\ B_\mu
\ \,&=&\, \ \sqrt{\,g^2\,+\,g'^2\,}\ \,
\left(\,J^\mu_{\ \,Z}\, =\, 
J^\mu_{\ \,3}\,-\,\sin^2\theta\,J^\mu_{\ \,\hbox{\tiny{em}}}\,\right)\ Z_\mu      
\nonumber \\
&& \ \ \ \ \ \ \ \, +\ \ g\ \sin\,\theta\ \,
\left(\,J^\mu_{\ \,\hbox{\tiny{em}}}\,=\,J^\mu_{\ \,3}\,+\,
\hbox{$\frac{1}{2}$} \ J^\mu_{\ \,Y}\,\right) \ A_\mu\ \ .
\nonumber 
\eea
This leads to expressions (\ref{Q}) and (\ref{e}) 
of the charge operator $\ Q\,=\,T_3\,+\,\frac{Y}{2}\ $ 
and the elementary charge $\ e\,=\,g\,\sin \theta$,
~and to expression (\ref{jz}) of the weak neutral current $\,J_Z$.
},  
\linebreak 
\vbox{\vskip.1truecm}

\noindent
in terms of the weak isospin and electromagnetic currents, by
\be
\label{jz}
J_Z\ \ =\ \ J_3\ -\ \sin^2 \theta \ J_{\footnotesize\hbox{em}}\ \ .
\ee
This shows the crucial r\^ole played, in the framework of the standard 
model and its neutral current phenomenology, ~by the electroweak mixing angle
$\,\theta$ and the associated $\,\sin^2 \theta$.

\vskip.2truecm

The new weak neutral current $\,J_Z\,$ should then manifest itself 
through a new type of neutrino-scattering processes, such as 
$\ \,\stackrel{\hbox{\tiny(\,$-$\,)}}{\nu}_\mu\,+ \ e^-\ \to \ \ 
\stackrel{\hbox{\tiny(\,$-$\,)}}{\nu}_\mu\,+ \ e^-\ $, 
~or 
$\ \,\stackrel{\hbox{\tiny($\,-\,$)}}{\nu}_\mu\,+ \ \,
\hbox{nucleon}\ \to \ \ \stackrel{\hbox{\tiny($\,-\,$)}}{\nu}_\mu\,+ \ X\,$, 
~which were effectively discovered~\cite{cn} 
at CERN and Fermilab, in 1973-1974.
~The $\,W^\pm\,$ and $\,Z\,$ bosons themselves could be directly produced in 
$ \,p\,\bar p\,$ collisions~\cite{z} (at CERN in 1983, and later at Fermilab)
and $\,e^+ e^- \,$ annihilations (at LEP and SLC),
providing a further confirmation of the ideas of gauge theories
through the direct observation of the massive gauge bosons.
\vskip .2truecm

\vbox{
Remember, however, that one does not only have to determine the value of 
$\,\sin^2 \theta\,$ in the standard model, despite the great importance 
of this parameter.
It was necessary, in the first place, to establish 
the validity of this model. Now that this is done (to a very large extent)
and that the corresponding value of $\,\sin^2\,\theta\,$ ($\,\simeq .231\,$)
is precisely known, it is essential to pursue 
the analysis of the radiative corrections, 
in connection with the understanding of 
the electroweak symmetry breaking mechanism and the searches for spin-0 
Higgs bosons.
And, at the same time,
to look for possible deviations which might signal the existence 
of ``new physics'' beyond the standard model. 
For all these purposes it is crucial to have
different types of experiments, performed at different energies, 
to study the various aspects of neutral-current interactions,
not only in neutrino scatterings, but also in electron interactions 
with matter.
\vskip .5truecm}

\section{Neutral currents in electron-hadron interactions}

\vskip .1truecm

\subsection{Neutral-current effects and parity violation 
in electron-hadron interactions}

Neutral current effects should be present in the interactions of electrons 
with matter, but are then in competition 
with ordinary electromagnetic interactions, 
which are normally much larger, as long as the center-of-mass 
energies and momenta remain small compared to the $\,W^\pm\,$ 
or $\,Z\,$ masses.
Indeed $\,\gamma $-exchange electromagnetic amplitudes are proportional to
$\,e^2/q^2$, \,while $Z $-exchange weak-
neutral-current amplitudes 
are proportional to $ \ G_F \,\simeq \,10^{-5}\ \,\hbox{GeV}^{- 2}$,
~with the $\,W^\pm\,$ and $\,Z\,$ masses 
\,-- which verify, at the classical level, 
$\,m_Z\,=\,m_W/\cos \theta$ --\, related to $\,G_F\,$ by 
\be
\label{mwmz}
\frac{G_F}{\sqrt 2}\ =\ \frac {g^2}{8\ m_W^{\ 2}}\ =\ 
\frac {g^2 + \,g'^2}{8\ m_Z^{\ 2}}\ \ .
\ee

\vskip .2truecm

Weak-interaction amplitudes, very small at low
energies owing to the very small value of the Fermi constant $\,G_F$,
\,become comparable in magnitude with electromagnetic amplitudes, for 
large values of the center-of-mass energies and momenta of the order of 
the $\,W^\pm\,$ and $\,Z\,$ masses.
At low or even mo\-de\-ra\-tely-high energies, however, 
the large effects of electromagnetic interactions could well prevent one 
to detect the effects of the weak-neutral-current
interactions between electrons and hadrons.
For example neutral-current amplitudes, proportional to $\,G_F$, ~are
very roughly of the order of $\ 10^{-4}\,$ times electromagnetic amplitudes, 
for values of the momentum transfer $\ |\,q^2|\,\approx \,1\ (\hbox{GeV}/c)^2$.

\vskip.2truecm

But weak interactions violate the discrete symmetry of parity, 
which exchanges the left and right orientations in space.
This is reflected by the fact that the charged weak currents 
to which the $\,W^\pm\,$'s are coupled 
are not pure vector currents as the electromagnetic current, 
but are given by a $\ V - A $ combination of vector and axial parts. 
This also corresponds to the fact that left-handed quark and lepton fields 
are associated in $\,SU(2)\,$ doublets, as indicated 
in Table \ref{tab:leptonq}, while their right-handed counterparts 
are $\,SU(2)\,$ singlets.
The $\,SU(2) \times U(1)\,$ structure of the electroweak theory 
requires that the new weak neutral current coupled to the $\,Z\,$ boson, 
given by $\ J_Z\, =\,J_3\ -\ \sin^2 \theta \ J_{\footnotesize\hbox{em}}\,$,
~must also violate parity.

\vskip .2truecm

It is then possible to take advantage of the fact that the 
interference terms between electromagnetic and weak-neutral-current amplitudes 
in electron-ha\-dron interactions violate parity,
to be able to detect the effects of the latter at moderately high energies;
and even also, in a more surprising way, at very low energies\,!
Actually parity-violation effects in electron-hadron interactions 
had already been considered by Zel'dovich~\cite{zel}, 
as early as in 1958, before the construction of the standard model.
Such interference effects between weak and electromagnetic amplitudes 
were indeed observed in a famous SLAC experiment~\cite{slac} in 1978,
in which a small asymmetry ($\,\approx \,10^{-\,4}\,$) 
in the deep-inelastic scattering 
of polarized electrons on deuterium could be measured.

\vskip .2truecm

Furthermore, neutral-current interactions can also induce mixings 
between atomic levels of different parities, an effect that 
is enhanced in heavy atoms and may then become accessible 
to observations, as discussed by M.A. and C. Bouchiat~\cite{bou} 
in their well-known paper of 1974.
These parity-violation effects in atomic physics, still very tiny, 
could indeed be detected and are now precisely measured 
by several experiments performed with 
different heavy atoms, cesium, bismuth, lead and thallium.

\vskip.2truecm

Both electron-hadron-scattering and atomic physics experiments 
give complementary informations on the four
coefficients 
$\,c_{1u}\,$ and $\,c_{2u}\,$, $\,c_{1d}\,$ and $\,c_{2d}\,$, 
~generally used to parametrize the parity-violating part of 
the effective Lagrangian density describing the neutral-current interactions 
of the electrons with the $\,u\,$ and $\,d\,$ 
quarks, main constituents of the protons and neutrons.
The parity-violating part of this low-energy effective 
Lagrangian density may be written\,\footnote{ See also, however, 
subsection \ref{subsec:lightgauge}, if relatively light neutral gauge bosons 
are present, in addition to the heavy $\,Z\,$ boson. 
Then the $\,c_{iq}'s\,$ don't have to be the same for different experiments
performed at different values of $\,q^2$\,!
Light $\,U\,$ bosons with parity-violating interactions were initially
discussed in supersymmetric theories with an extra 
$\,U(1)\,$ gauge group, but their possible existence should be 
considered independently of this original motivation. 
\medskip} as follows:
\be
\label{defc}
{\cal L}_{\footnotesize\hbox{eff}}\ \ =\ \ \frac{G_F}{\sqrt 2}\ \ \,
\Sigma_{q = u,d}\ \left[\ \
c_{1q}\ \ \,\bar q\ \gamma^\mu\,q\ \ \,\bar e \ \gamma_\mu\gamma_5\,e\ +\ 
c_{2q}\ \ \,\bar q\ \gamma^\mu\gamma_5\,q\ \ \,\bar e \ \gamma_\mu\,e\ \ 
\right]\ \ .
\ee

\subsection{Parity-violation in electron-hadron scattering experiments}

The weak-neutral-current amplitudes corresponding 
to this effective Lagrangian density interfere 
with ordinary electromagnetic amplitudes, 
induced by the exchanges of a photon between 
the $\,u\,$ or $\,d\,$ quarks 
(of charges $\,\frac{2}{3}\,$ and $\ -\,\frac{1}{3}\,$) and the electron.
For the SLAC experiment~\cite{slac} with an isoscalar deuterium target,
\be
e^-_{\ R,L}\ +\ \hbox{deuterium}\ \ \to\ \ e^-\ +\ X\ \ ,
\ee
performed with polarized electrons of energies $\,\approx$ 20 GeV,
and momentum transfer
$\,Q^2\,=\,-\,q^2\,$ (defined to be positive) 
of the order of $\ \,\approx\,1.6\ (\hbox{GeV}/c)^2$,
~the interference term in the deep-inelastic scattering cross-section 
on $\,u\,$ and $\,d\,$ quarks 
is a linear combination of the two quantities 
$\ (\,\frac{2}{3}\,) \ c_{iu}\ +\ (\,-\,\frac{1}{3}\,)\ c_{id}\,$, ~i.e.
\be
\label{combc}
c_{1u}\ -\ \hbox{$\frac{1}{2}$}\ \,c_{1d}\ \ \ \ \ \ \hbox{and}\ \ \ \ \ \ 
c_{2u}\ -\ \hbox{$\frac{1}{2}$}\ \,c_{2d}\ \ \ .
\ee

\vskip .1truecm
We can easily compute an approximate value of the asymmetry 
by neglecting the contribution of the $\,c_{2q}$'s (which are 
small in the standard model, 
for $\,\sin^2\theta \simeq \frac{1}{4}\,$) \footnote{
In this approximation 
both electromagnetic and weak amplitudes for the scattering of electrons 
on $\,u\,$ and $\,d\,$ quarks involve the quark vector current 
$\,\bar q\, \gamma^\mu q\,$.
~The polarized cross-sections may be written as:
\bea
\sigma_{R,L}\ \ \propto\ \ 
\left[\ \frac{2}{3}\ \frac{-\ e^2}{q^2}\ \pm\ 
\frac{G_F}{\sqrt2}\ c_{1u} \ \right]^2\ \ +\ \ 
\left[\ -\ \frac{1}{3}\ \frac{-\ e^2}{q^2}\ \pm\ 
\frac{G_F}{\sqrt2}\ c_{1d} \ \right]^2\ \ . \nonumber
\eea
The interference terms between electromagnetic and weak amplitudes 
generate an asymmetry
\bea
A\ \ =\ \ \frac{\sigma_R\,-\ \sigma_L}{\sigma_R\,+\ \sigma_L}\ \ =\ \ 
2\ \,\frac{G_F}{\sqrt 2}\ \,\frac{Q^2}{e^2}\ \ 
\frac{\frac{2}{3}\ c_{1u}\ -\ \frac{1}{3}\ c_{1d}}
{(\frac{2}{3})^2\,+\,(-\frac{1}{3})^2}\ \ =\ \  
\frac{3\ G_F\ \,Q^2}{5\ \sqrt 2\ \pi \,\alpha}\ \
(\,c_{1u}\,-\,\hbox{$\frac{1}{2}$}\ c_{1d}\,)\ \ .
\nonumber
\eea
With $\ \frac{3\,G_F}{5\ \sqrt 2\,\pi \,\alpha} \,\simeq\,2.16\ \,10^{-\,4}\ \,
(\hbox{GeV})^{-\,2}$, ~and 
$\ (\,c_{1u}\,-\,\frac{1}{2}\ c_{1d}\,)\,\simeq\, -\,.36\,$ 
in the standard model, we get an approximate value of 
$\ A/Q^2 \,\simeq \,- \ 8\,\ 10^{-\,5}\ (\hbox{GeV}/c)^{-\,2}$.
\medskip \\
}. 
~More precisely, the asymmetry\,\footnote{ 
The polarized cross-sections on $\,u\,$ and $\,d\,$ quarks may be written as:
\bea
\ba{ccc}
\sigma_{R,L}^{\ u} \ &\propto& \
\left[\,\frac{2}{3}\ \frac{-\ e^2}{q^2} \pm
\frac{G_F}{\sqrt2}\ (c_{1u}+ c_{2u}) \,\right]^2 \, + \
\left[\,\frac{2}{3}\ \frac{-\ e^2}{q^2} \pm 
\frac{G_F}{\sqrt2}\ (c_{1u}- c_{2u}) \,\right]^2\,(1-y)^2\ ,\vspace{3mm}\\
\sigma_{R,L}^{\ d} \ &\propto& \
\left[-\frac{1}{3}\ \frac{-\,e^2}{q^2} \pm
\frac{G_F}{\sqrt2}\ (c_{1d}+ c_{2d}) \right]^2  \,+ \
\left[-\frac{1}{3}\ \frac{-\,e^2}{q^2} \pm 
\frac{G_F}{\sqrt2}\ (c_{1d}- c_{2d}) \right]^2 \ (1-y)^2\  .
\ea  \nonumber 
\eea
The interference terms between electromagnetic and weak amplitudes 
generate an asymmetry
\tiny
\bea
A\ =\ 2\ \,\frac{G_F}{\sqrt 2}\ \,\frac{Q^2}{e^2}\ \ 
\frac{\frac{2}{3}\ 
\left[\,(c_{1u}+ c_{2u})\,+\,(c_{1u}- c_{2u})\,(1-y)^2\,\right]\, 
-\,\frac{1}{3}\,\left[\,(c_{1d}+ c_{2d})\,+\,(c_{1d}- c_{2d})\,(1-y)^2\,\right]}
{\left[\,(\frac{2}{3})^2\ +\ (-\frac{1}{3})^2\,\right]\ 
\left[\,1+(1-y)^2\,\right]} \ \,.
\nonumber 
\eea
\footnotesize
This leads directly to expression (\ref{asym}).
}
$\ A\,=\,\frac{\sigma_R\ -\ \sigma_L}{\sigma_R\ +\ \sigma_L}\,$,
~proportional to 
{\Large $\ \frac{G_F}{\alpha/q^2}$}\,, 
~is given by:
\be
\label{asym}
\frac{A}{Q^2}\ \ =\ \ \frac{3\ \,G_F}{5\ \sqrt 2\ \pi \,\alpha}\ 
\left[\ (\,c_{1u}\,-\,\hbox{$\frac{1}{2}$}\ c_{1d}\,)\ +\ 
(\,c_{2u}\,-\,\hbox{$\frac{1}{2}$}\ c_{2d}\,)\ \,
\frac{1\,-\,(1-y)^2}{1\,+\,(1-y)^2}\ \,\right]\ \ ,
\ee
in which $\,y\,$ (\,$\sim .2\,$) is the fraction of the incoming electron 
energy transferred to the hadrons in the final state.
The measured value of the asymmetry, given by 
$\ A/Q^2 \,=\, (\,- \ 9.5 \,\pm \,1.6\,)\ 10^{-\,5}\ (\hbox{GeV}/c)^{-\,2}$,
has been used to discrimate the standard model from a number of possible 
alternatives. 
With a further analysis of the $\,y\,$ dependence 
of the asymmetry (\ref{asym}), 
these experiments provided combined constraints on the two quantities
$ \ (\,c_{1u} - \frac{1}{2}\ c_{1d}\,)\,$ 
and $ \ (\,c_{2u} - \frac{1}{2}\ c_{2d}\,)\,$.
~And also, in the framework of the standard model
(for which the $\,c_{iq}$'s are known functions of $\,\sin^2 \theta$,
~given in subsection \ref{sec:ncsm}),
a determination of  $\ \sin^2\theta \ \simeq \,.224\,\pm \,.020\,$. 
\\
\vskip .1truecm

We refer, more generally, to the following lectures~\cite{keo}
for a description of \linebreak
the different electron-hadron scattering experiments
that have been performed.

{\vbox{\vskip 0truecm}

\subsection{Parity-violation in atomic physics experiments}

In atomic physics, for which a typical momentum transfer may be taken as
$\,q\,\sim\,m_e c\,\alpha\,$,
~one might na\"{\i}vely anticipate parity-violation effects 
to be of the order of 
{\large $\ \frac{G_F}{\alpha/q^2}$}\,, 
~leading to rough expectations 
$\ \sim\, G_F\,m_e^{\,2}\ \alpha \,\sim \,10^{-14}$,
considerably smaller than the 
$\,\sim 10^{-4}\,$ at $\,|\,q^2|\,\sim \,1 \ (\hbox{GeV}/c)^2\,$ that we just 
discussed for the SLAC experiment.

\vskip .2truecm
For a heavy atom however~\cite{bou}, the momentum transfer corresponding 
to a penetrating electron in the vicinity of a nucleus of charge $\,Z\,$ 
is $\ q\,\sim\, m_e c\ Z\,\alpha$, ~and this electron interacts 
coherently with the constituents of the nucleus. 
As a result the previous estimate, now replaced by 
{\large $\ \frac{Z\,G_F}{\alpha/q^2}$}
$\, \sim \, G_F\,m_e^{\,2}\ \alpha \ Z^3$,
~is enhanced by a factor which behaves roughly like $\,Z^3\,$.

\vskip .2truecm

More precisely, parity-violation effects in atomic physics 
are mostly sensitive to the {\it \,vector\,} part of 
the quark weak-neutral-current,
combined with the {\it \,axial\,} part of the electronic current, 
which can induce mixings between atomic levels of different parities.
This term involving the vector part of the quark weak-neutral-current 
in the effective Lagrangian density (\ref{defc}) 
is parametrized by the two coefficients 
$\,c_{1u}\,$ and $\,c_{1d}\,$, 
~while a nucleus with $\,Z\,$ protons and $\,N\,$ neutrons 
includes $\,(2\, Z + N)\,$ $\,u\,$ quarks and $\,(Z + 2\, N)\,$ $\,d\,$ quarks.

\vskip .2truecm

In the non-relativistic limit for which the electron may be described 
by a two-component spinor $\,\varphi$ 
~(with small components expressed as 
$\,\chi\,\simeq\,\frac{\vec \sigma . \,\vec p}{2\ m_e}\ \varphi\,$), 
~the relevant part in the parity-violating Lagrangian density 
(\ref{defc}) reads:
\be
{\cal L}_{\hbox{\tiny{P.V.}}}^{\ (1)}\ \ \simeq\ \ \frac{G_F}{\sqrt 2}\ \ 
\left[ \ (2\, Z + N)\ c_{1u}\,+\,(Z + 2\, N)\ c_{1d}\ \right]\ \
\delta^3(\vec r)\ \ 
(\ \varphi^\dagger\ \frac{\vec \sigma . \,\vec p}{2\ m_e}\ \varphi
\ \ +\ \ \hbox{c.c.}\ )\  .
\ee
In terms of the ``weak charge'' of the nucleus, defined as 
\be
\label{qw}
Q_W\ \ =\ \ -\ 2\ 
\left[ \ c_{1u}\ (2\, Z + N)\ +\ c_{1d}\ (Z + 2\, N)\ \right]\ \ ,
\ee
this corresponds to the following effective parity-violating potential 
between an electron and a nucleus:
\be
\label{pv}
V_{\hbox{\tiny{P.V.}}}^{\ (1)}\ \ \simeq\ \ 
\frac{G_F}{4\,\sqrt 2}\ \ \frac{Q_W}{m_e}\ \ 
\left(\ \vec \sigma . \,\vec p\ \,\delta^3(\vec r)\ +\ 
\delta^3(\vec r)\ \,\vec \sigma . \,\vec p\ \right)\ \ .
\ee
(The other part in the effective Lagrangian density (\ref{defc}), 
parametrized by the $\,c_{2q}\,$ coefficients, involves the axial 
quark currents and leads to additional parity-violating 
contributions, which are proportional to the spin of the nucleus.)

\vskip .2truecm

Due to the very small value of the Fermi constant (corresponding to 
$\,G_F\,m_e^{\ 2} \,\alpha\,\simeq \ 2\ \,10^{-14}\,$) 
~the parity-violation effects
might have been totally negligible at low energies.
But as shown by the Bouchiat's they are significantly enhanced in heavy atoms, 
since the matrix elements of $\ V_{\hbox{\tiny{P.V.}}}^{\ (1)}$
between two $\,S\,$ and $\,P\,$ states mixed by this parity-violating 
interaction 
involve the product of a $\,s\,$ wave function near the origin
(which behaves like $\,\sqrt Z\,$), 
times the derivative of a $\,p\,$ wave function 
(which behaves like $\,Z \,\sqrt Z\,$), 
~times the weak charge $\,Q_W\,$ (which behaves like $\,Z$).

\vskip .2truecm 

Owing to the resulting $\ \approx\,Z^3$ enhancement factor,
in particular, 
these parity-violation effects become accessible to experiments,
and have now been measured, for several heavy atoms, 
at a level of precision that can be better that 1\,\%\,~\cite{bou}.
The results are well in agreement with the values expected from 
standard model calculations.
In the case of cesium ($Z = 55\,$ with $\,Z+N =133$) for example,
the measured and ``standard model'' 
values of the weak charge~\cite{mesqw,lan}
are given below, and verify:
\be
\label{difqw}
Q_{W\,\hbox{\tiny{exp}}}\,-\,Q_{W\,\hbox{\tiny{SM}}}\,=\,
(\,-\,72.4\,\pm\, 0.3_{\hbox{\tiny{exp}}}\,\pm\, 0.8_{\hbox{\tiny{th}}}\,)\,
-\,(\,-\,73.1\,\pm\, 0.1\,)\,\simeq\, 0.7\,\pm\,0.9\ .
\ee

\vskip .2truecm

For heavy atoms expression (\ref{qw}) of $\,Q_W\,$ 
corresponds to a linear combination of
$\,c_{1u}\,$ and $\,c_{1d}\,$ that is quite different 
(and very roughly orthogonal) to the linear combination 
$\ (\,c_{1u}\,-\,\frac{1}{2}\ c_{1d}\,)\,$ 
to which the SLAC experiment is sensitive.
The two types of experiments then play complementary r\^oles
in the determination of the exact structure of the electron-hadron
weak-neutral-current interactions.

\subsection{Parity-violation and the neutral current in the Standard Model}

\label{sec:ncsm}

In the standard model, there is a single weak neutral current, given by 
$\ J_Z\,=\,J_3\ -\ \sin^2 \theta \ J_{\footnotesize\hbox{em}}$.
~The corresponding contributions of the electron, 
and the $\,u\,$ and $\,d\,$ quarks, to this weak neutral current, read 
as follows:
\be
(J^\mu_{\ Z})_e\ \ =\ \ 
(\,-\ \hbox{$\frac{1}{4}$}\ +\ \sin^2\theta\,)\ \ \bar e \ \gamma^\mu\,e\ +\ 
\hbox{$\frac{1}{4}$}\ \ \bar e \ \gamma^\mu\gamma_5\,e\ \ ,
\ee
for the electronic weak neutral current, and
\be
\left\{ \ \
\ba{ccccc}
(J^\mu_{\ Z})_u\ \,&=&\ \ 
(\,+\ \frac{1}{4}\ -\ \frac{2}{3}\ \sin^2\theta\,)\ \ \bar u \ \gamma^\mu\,u
\ &-&\ \frac{1}{4}\ \ \bar u \ \gamma^\mu\gamma_5\,u\ \ ,   \vspace{3mm}\\
(J^\mu_{\ Z})_d\ \,&=&\ \ 
(\,-\ \frac{1}{4}\ +\ \frac{1}{3}\ \sin^2\theta\,)\ \ \bar d \ \gamma^\mu\,d  
\ &+&\ \frac{1}{4}\ \  \bar d \ \gamma^\mu\gamma_5\,d\ \ ,
\ea  \right.
\ee
for the $\,u\,$ and $\,d\,$ quark weak neutral currents.

\vskip .2truecm

{From} the effective Lagrangian density
\be
{\cal L}_{\hbox{\tiny{eff}}}\ \ =\ \ 
-\ \,\frac{g^2\,+\,g'^2}{m_Z^{\ 2}}\ \,(J^\mu_{\ Z})_q\ (J_{\mu\ Z})_e\ \ =\ \ 
-\ \,8\ \,\frac{G_F}{\sqrt 2}\ \,(J^\mu_{\ Z})_q\ (J_{\mu\ Z})_e\ \ ,
\ee
we get the expressions (at the classical level) of the four coefficients 
$\,c_{iu}$ and $\,c_{id}\,$ 
which parametrize the effective Lagrangian density (\ref{defc}): 
\be
\left\{\ \ 
\ba{ccrcr}
c_{1u}\ \,&\simeq&\ -\ \frac{1}{2} \,&+&\,\frac{4}{3}\ \sin^2\,\theta\ \ ,
\vspace{2mm} \\
c_{1d}\ \,&\simeq&\ \ \ \ \ \frac{1}{2} \,&-&\, \frac{2}{3}\ \sin^2\,\theta
\ \ , \vspace{4mm} \\
c_{2u}\ \,&\simeq&\ -\ \frac{1}{2}\,&+& \,2\ \sin^2\,\theta \ \ , 
\vspace{2mm} \\
c_{2d}\ \,&\simeq&\ \ \ \ \ \frac{1}{2} \,&-&\, 2\ \sin^2\,\theta\ \ . 
\vspace{1mm}\\
\ea  \right.
\ee
And the weak charge $\,Q_W\,$ defined by eq.\,(\ref{qw}) is given
(again up to small higher-order corrections)
by the following expression\,\footnote{ This may also be obtained directly
in the standard model,
from the effective Lagrangian density
\bea
{\cal L}_{\hbox{\tiny{P.V.}}}^{\ (1)}\ \ =\ \ 
-\ \,2\ \ \frac{G_F}{\sqrt 2}\ \ 
(J^\mu_{\ Z})_{q\,\hbox{\tiny{vect.}}}\ \ \bar e \ \gamma_\mu\gamma_5\,e\ \ .
\nonumber
\eea
The vector part in the hadronic $\,Z\,$ current leads to a density
$\ [\,\frac{1}{4}\ (Z-N)\ -\ \sin^2\theta\ Z\,]\ \delta^3(\vec r)\,$,
and therefore to the parity-violating potential (\ref{pv}), with 
expression (\ref{qw2}) of the weak charge $\,Q_W$.
}, which depends on $\ \sin^2 \theta\,$:
\be
\label{qw2}
Q_W\ \ \simeq\ \ Z\ \,(\,1\,-\,4\,\sin^2\,\theta\,)\ -\ N\ \ .
\ee

\vskip .1truecm

To fix ideas, a rough evaluation of the $\,c_{iq}\,$  parameters
would be, for $\ \sin^2\theta\,\simeq .225$, 
$\ c_{1u}\,\simeq\, -\ .20,\ c_{1d}\,\simeq \, .35,\ 
c_{2u}\,\simeq \, -\ .05,\ c_{2d}\,\simeq \,.05\,$.
And the weak charge would then be given by 
$\ Q_W\,\simeq\,.1\ Z - N\,\simeq\,-\,72.5\,$, ~in the case of cesium
(for which $\,Z=55$ ~and $\,N=78$), ~which practically coincides with
the measured value given in eq.\,(\ref{difqw}).
~But precise estimates require the consideration of radiative corrections.

\vskip .3truecm

Electron-hadron scattering and atomic physics experiments 
led, in the framework of the Standard Model, 
to precise determinations of the $\,\sin^2 \theta$ \linebreak parameter,
as it will be discussed in subsequent lectures~\cite{bou,keo,musolf,altarelli}.
Maybe even more importantly, they played a crucial r\^ole in discriminating 
this model from possible alternatives, 
through the determination of the exact structure of the weak neutral current.
More precisely one had to establish whether all experiments involving 
weak-neutral-current effects, including of course neutrino and $\,e^+\,e^-$ 
scattering experiments, could be interpreted, or not,
in terms of a single weak neutral current; and if so, if this neutral current 
did actually obey eq.\,(\ref{jz}), 
$J_Z\,=\,J_3\ -\ \sin^2 \theta \ J_{\footnotesize\hbox{em}}\,$,
~and for which value of $\ \sin^2 \theta\,$.

\vskip .2truecm
Up to now only one weak neutral current $\,J_Z\,$, ~obeying eq.\,(\ref{jz}),
was exhibited experimentally. 
And the experimental determinations of the weak-neutral-current 
interaction parameters~\cite{lan}
\be
\begin{tabular}{|c|c|c|}
\hline  && \\
&  experiment & standard model  \\ [.3 true cm] \hline 
&& \\ 
$c_{1u}$   &  $-\, .216\,\pm \, .046 $ &  $ - \, .188 5\,\pm \, .000 3 $  
\\  [.3 true cm] 
$c_{1d}$   &  $ \phantom{-}\,.361\,\pm \, .041  $  &   $ 
\phantom{-}\, .3412\,\pm \, .0002$  
\\  [.3 true cm] 
$\ c_{2u}  \ -\ \frac{1}{2}\ \, c_{2d}\ $    &   $ - \ .03\ \pm \  .12$   &   
$-\, .0488\,\pm \, .0008$   
\\  [.3 true cm] \hline  && \\ 
$Q_W$\,(Cs) & $-\,72.41\,\pm \,.25\,\pm \,.80 $&$ -\,73.12\,\pm\,.06 $
\\  [.3 true cm]
$Q_W$\,(Tl) & $-\,114.8\,\pm \,1.2\,\pm \,3.4 $&$ -\,116.7\,\pm\,.1\,\ $ 
\\  [.3 true cm]
\hline
\end{tabular}
\ee
are in good agreement with the standard model values (here evaluated for 
$\,m_{\hbox{\tiny{Higgs}}} \,\simeq\,m_Z$). One has also, in addition, 
$\,c_{2u}\,+\,c_{2d}\,\simeq \, -\, .0095$, in the standard model.

\vskip .4truecm

On the other hand many theoretical ideas involve additional 
weak neutral currents coupled to extra neutral gauge bosons. 
The possible existence of such additional gauge bosons is frequently suggested, 
for example in left-right symmetric or grand-unified theories, 
and in theories with extra $\,U(1)\,$ gauge groups, 
often motivated by supersymmetry or superstring ideas.
These ideas of left-right symmetry 
(with gauge group
$\,SU(3) \times SU(2)_L \times SU(2)_R \times U(1)_{B-L}\,$)
and grand-unification 
(with larger gauge groups such as $\,O(10)\,$ or $\,E(6)\,$)
also provide natural frameworks in which neutrinos could have
tiny masses and oscillate from one flavor to another.

\vskip.2truecm
It is worth to emphasize, at this stage, that experiments working 
at low values of the momentum transfer 
\,-- and in particular atomic physics experiments --\,
could be sensitive 
to neutral gauge bosons ($\,U\,$)
that would be both relatively light and very weakly coupled,
thereby escaping detection in experiments performed 
higher $\,|\,q^2|\,$ large compared to $\,m_U^{\ 2}$.
~We shall return to this point in subsection \ref{subsec:lightgauge}.

\vbox{\vskip .4truecm}

\section{Electroweak physics and precision tests of the Standard Model}
\label{sec:electroweak}

\vskip .2truecm

Altogether many experimental data have confirmed the validity of the 
Standard Model (although we still have no detailed experimental
information about the mechanism by which the $\,SU(2) \times U(1)\,$ 
electroweak symmetry gets broken).
Let us mention, in particular, the direct production 
of the $\,W^\pm\,$ and $\,Z\,$ bosons at the CERN and Fermilab $\,p\,\bar p\,$ 
colliders, the detailed study of the properties of the $\,Z\,$ boson 
at LEP 1 and SLC, 
and, more recently, the direct production of $\ W^\pm\,$ pairs at LEP 2.
Also the top quark, necessary for the consistency of the theory,
was found at Fermilab with a mass $\,m_{\hbox{\tiny{top}}}\,\simeq\,170 - 180 
\ \hbox{GeV}/c^2$, ~in agreement with the expectations derived from the 
analysis of the radiative corrections.

\vskip .2truecm

The standard model of strong, electromagnetic and weak interactions 
gives, at the present time, a remarkably good description of the interactions 
of particles, very well understood from the exchanges of spin-1 gauge bosons
between spin-$1/2$ leptons and quarks\,\footnote{ We leave
aside the question of neutrino masses and oscillations, 
two phenomena which are absent from the standard model defined 
{\it stricto sensu}, but may be incorporated easily 
through the introduction of right-handed neutrino fields.}.
The consistency of all experimental results with
this standard model, for a value of the electroweak  mixing angle
given by $\,\sin ^{2}\theta \,\simeq\,.231$, 
~now requires that one takes into account the effects of electroweak 
radiative corrections. And experiments are getting sensitive to the effects 
of spin-0 Higgs bosons associated with the spontaneous breaking 
of the electroweak symmetry, a point to which we shall return later. 
The running of the strong interaction coupling constant $\,\alpha _{s}$,
measured up to LEP energies ($\,\alpha _{s}(m_{Z})\,\simeq\,.12$), \,is also 
in good agreement with QCD expectations.

\vskip .2truecm

The precise determination of these numbers, 
including the electromagnetic constant $\,\alpha(m_Z)\,\simeq \,1/{129}$,
~is essential in the discussion of a possible unification 
of the three (suitably-normalized) 
$\,SU(3)\,$, $\,SU(2)\,$ and $\,U(1)\,$ gauge couplings at high energies, 
as expected in grand-unified theories. 
As we shall see in section \ref{sec:symgut}, 
such an unification does not occur if the evolution 
of the couplings from ``low'' energies (i.e. $\approx\, m_Z$) 
to very-high energies 
is computed with the particle spectrum of the standard model.
On the other hand a unification of the values of the three gauge couplings 
is obtained, at an energy of the order of $\,10^{16}$ GeV$/c^2\,$ or so,
~if the evolution of the three gauge couplings is governed by the 
particle spectrum of the Supersymmetric Standard Model.
This is often taken as a possible (indirect) indication 
in favor of supersymmetry. 
Before discussing briefly such extensions of the Standard Model, let us 
say a few things about Higgs bosons.

\vskip .2truecm

In the Standard Model the spontaneous breaking of the electroweak gauge 
symmetry is induced by fundamental spin-0 fields known as (Englert-Brout) 
Higgs fields~\cite{sm1}$^-$\,\cite{higgs}.
This requires the existence of a new neutral spin 0 particle called the 
Higgs boson. 
This particular aspect of the Standard Model has not been confirmed 
experimentally yet, and one can still question the actual existence 
of such a particle, which remains the only missing ingredient 
of the model, after the experimental discovery of the top quark.

\vskip .2truecm

Since the Higgs boson has escaped all direct searches performed 
at LEP up to now, its mass must be larger than about 
80 GeV$/c^{2}$, a limit that recently increased up to about 
90 GeV$/c^{2}\,$ and even more, 
with the increase in the energies accessible at LEP 2. 
Can the mass of this particle, essential for the renormalisability 
of the theory, be very large\,? 
Then we could no longer think of the Higgs boson as an 
ordinary elementary particle: one has
$\,\Gamma_{\hbox{\tiny{Higgs}}}\,\approx \,\frac{1}{2}\ m_H^{\ \,3}$, 
~if the Higgs mass and width are measured in TeV's,
~so that a \,1 TeV$/c^2\,$ Higgs would have a width of about 
$\,\frac{1}{2}\,$ TeV$/c^2\,$!

\vskip .2truecm

Moreover the coupling constant $\,\lambda\,$ which governs the magnitude 
of the Higgs boson self-interactions grows like the square 
of the Higgs mass\,\footnote{ \label{pothiggs} More precisely, 
if we write the potential of the spin-0 doublet Higgs field $\,\varphi\,$ as
\bea
V(\,\varphi\,)\ \ =\ \ \lambda\ (\,\varphi^\dagger \varphi\,)^2\ -\ 
\mu^2\ \varphi^\dagger \varphi\ \ ,
\nonumber
\eea
the neutral component of the Higgs field acquires a non-vanishing 
vacuum expectation value $\ < \!\varphi^0\!>\ \,=\,\frac{v}{\sqrt 2}\,$, 
\,~with 
$\ \,v\,=\,\sqrt{\,\frac{\mu^2}{\lambda}}\,$.
The $\,W^\pm\,$ and $\,Z\,$ masses are $\ m_W \,=\, g\,v/2\,$, 
$\ m_Z \,=\, \sqrt{\,g^2\,+\,g'^2} \ \,v/2\,$,
~\,so that 
\bea
\frac{G_F}{\sqrt 2}\ \ =\ \ \frac{g^2}{8\,m_W^{\ 2}}\ \ =\ \ 
\frac{g^2\,+\,g'^2}{8\,m_Z^{\ 2}}\ \ =\ \ \frac{1}{2\ v^2}\ \ ,
\nonumber
\eea
which determines 
$\,\ v\,=\,(\,G_F\,\sqrt 2\ )^{-\,\frac{1}{2}}\,\simeq \,246\ \,$GeV.
~The Higgs mass being given by $\ m_H^{\ 2}\,=\,2\ \mu^2$, \,~one can
express the Higgs self-coupling constant as
\bea
\lambda\ \ =\ \ \frac{\mu^2}{v^2}\ \ =\ \ 
\frac{G_F}{\sqrt 2}\ \ m_H^{\ \,2}\ \ .
\nonumber
\eea
In particular $\ \lambda ^2 /4\pi\,\simle\,1$, ~for 
$\ m_H\ \simle\ \pi^{\frac{1}{4}}\ 2\ v\ \simeq\ 650\,$ GeV.
},
so that, if we want to be able to compute perturbatively 
up to energies of the order of the Higgs mass at least, 
this mass should be less than about 800 GeV$/c^2$ or so.
(If, furthermore, we demand to be able to compute perturbatively up to a large 
grand-unification scale of the order of $\,10^{16}\,$ GeV/$c^2$, 
~the Higgs mass should be less than about 200 GeV/$c^2$; 
~it should also, at the same time, be larger than about 120 GeV/$c^2$,
if the standard model vacuum state considered is to remain stable
or at least metastable.)

\vskip .2truecm

However, unlike in the case of the top quark, lowest-order
radiative corrections to the presently measured quantities 
are not very sensitive to the Higgs boson mass 
(their dependence in $m_H$ being generally logarithmic, 
rather than quadratic as in the case of $\,m_{\hbox{\tiny{top}}}$).
At the present stage, the analysis of radiative 
corrections indicates that the Higgs mass should be less than 
about \,300 \,GeV$/c^2$, ~in the framework of the standard model. 

\vskip .2truecm
\vbox{
Should one expect a future discovery of this elusive particle, 
and then consider it as the final
confirmation of an achieved and satisfactory theoretical construction\,?
Unfortunately not, since, despite the remarkable achievements of this 
model, it cannot be considered as a complete and satisfactory theory 
of the fundamental laws of Nature.
 Indeed the standard model 
leaves, on the theoretical side, many questions unanswered. 
\vskip .2truecm
}

\section{Open questions in the Standard Model}
\label{sec:openquestions}

\vskip .1truecm

\subsection{The origin of the charges, and the question of parity conservation}

Why is the electric charge quantized, and the quark charges 
$\,+\,2/3\,$ and $\,-\,1/3\,$ (times the elementary unit charge $\,e$)\,?
These values, given by $\,Q = T_3 + \frac{Y}{2}\,$,
follow from appropriate choices for the weak hypercharges 
$\,Y\,$ of  the chiral quark and lepton fields: $\,Y = -\,1\,$ 
for the left-handed electron field $\,e^-_{\ L}\,$, but $\,-\,2\,$ for 
$\,e^-_{\ R}\,$; 
$\,Y = 1/3\,$ for the quark fields $\,u_L\,$ and $\,d_L\,$, but
$\,4/3\ $ and $\, -\,2/3\, $ for 
$\,u_R\,$ and $\,d_R\,$. Why these rather special values\,?
And at first, why is parity violated in weak interactions 
(left-handed quark and lepton fields being members of $\,SU(2)\,$ 
doublets, and right-handed ones singlets),
while it remains conserved by electromagnetic ones\,?
This is well described by the standard model, but 
we would like to understand 
the origin of the $\,SU(3) \times SU(2) \times U(1)$ gauge 
symmetry properties of quark and lepton fields.

\subsection{The ``family problem'' and the origin of the lepton and quark mass 
spectrum}

Why are there three families of leptons and quarks 
$\,(\,\nu _{e},\ e,\ u,\ d\,$), \linebreak $(\,\,\nu _{\mu }$, $\mu $, $c$,
$s\,$), $\,(\,\nu _{\tau }$, $\tau $, $t$, $b$\,), rather than a single one, 
and how
could their mass  spectrum and mixing
angles be theoretically  understood\,? This is in fact an old question, 
which dates back to the discovery of the muon. Is there a symmetry 
between the three families, and if so, why and how gets it broken\,?
Are neutrinos exactly massless \,-- as in  the standard model 
defined {\it \,stricto sensu} --\, or slightly massive\,? 
In that case, do they ``oscillate'' 
from one flavor to another, $\,\nu_e$'s, for example, being transformed 
into $\,\nu_\mu$'s or $\,\nu_\tau$'s\,, and conversely\,? 
Could such oscillations be responsible for the 
deficit of the measured flux of neutrinos produced by the Sun, compared to
 expectations based on the standard model of the Sun\,? 
Could $\,\nu_\mu\ \to\ \nu_\tau\,$ oscillations, for example, 
 be responsible 
for the anomalies presented by the observed fluxes 
of atmospheric neutrinos~\cite{nu}\,?
Could massive neutrinos, or other particles, provide some 
of the ``dark matter''  that seems to be present in the Universe\,?

\subsection{The $\,CP\,$ problem, and the origin of matter in the Universe}

The $\,CP\,$ symmetry which relates matter to antimatter 
 -- while exchanging the left and right orientations in space  -- is a symmetry 
for almost but not all interactions, since it is
slightly violated in weak decays of neutral kaons.
This $\,CP\,$ violation is generally attributed to the effect of the 
(weak) $\,CP$-violating phase parameter $\,\delta\,$ in the 
Cabibbo-Kobayashi-Maskawa mixing matrix between the three quark families,
although this remains to be proven. 
Once $\,CP\,$ is broken it has no reason to be an exact 
symmetry of the strong interactions,
and the neutron should be expected to have some 
electric dipole moment. 
Since none has been found the corresponding amount of ``strong''
$\,CP$-violation, ~measured by the effective dimensionless parameter 
$\,\theta _{\rm {QCD}}\,$, should be smaller than $\,\approx 10^{-9}$. 
Why should $\,\theta _{\rm QCD}\,$ be so small\,?
 A possible explanation involves extensions of the standard 
model with a new broken $\,U(1)\,$ symmetry 
called the Peccei-Quinn symmetry.
This would require the existence of a neutral, very light, 
spin-0 Goldstone-like particle, the axion~\cite{axion}, 
decaying into photon pairs. 
But none has been observed, and 
its existence is rather constrained, 
both from particle physics experiments and astrophysical arguments.

\vskip .2truecm

 A related question concerns the density of matter 
in the Universe. Why is the average number of nucleons 
of the order of a few $10^{-\,10}\,$ 
compared to the number of photons in the primordial $\,2.7\,^\circ K$  cosmic 
microwave background radiation\,? What is the origin of these nucleons\,? 
Are they remnants of the annihilation of matter with antimatter
in the very early Universe, 
an extremely small excess of matter 
being present since the origin of time\,? 
If not, could this very small excess of matter 
have been generated from an initially symmetric Universe\,? 
As noted by Sakharov~\cite{sak} in 1966, this would require the existence of 
interactions that do not conserve the baryon number $\,B$,
so as to generate a net excess of nucleons; and do not conserve the 
charge conjugation $\,C\,$ and  $\,CP\,$ symmetries, 
so that matter and antimatter can evolve 
asymmetrically, through out of equilibrium phenomena. 
What could be these $\,B\,$, $\,C\,$ and $\,CP$-violating interactions\,? 
Perhaps those of a ``grand-unified'' theory relating quarks to leptons; 
or maybe those which 
should appear at very high energies within the standard model itself, 
when non-perturbative effects of electroweak interactions 
are taken into account\,?

\subsection{``Too many arbitrary parameters''}

The standard model depends in fact, not on a single parameter
$\,\sin^2\theta$, ~but on a total of {\it \,twenty\,} arbitrary parameters:
the three gauge coupling constants
$\ g_{s\,}$, $\,g\,$ and $\,g'$ ~of the 
$\,SU(3) \times SU(2) \times U(1)\,$ gauge group; 
the two parameters $\ \mu ^{ 2}\,$ and $\,\lambda\,$ 
which determine the mass 
and  self-interactions of the Higgs boson;
the nine quark and charged-lepton masses,
plus the three quark-mixing angles and the 
(weak) $\,CP$-violating phase $\,\delta$;
with also, in addition, 
the (strong) $\,CP$-violating parameter  $\,\theta _{\rm {QCD}}\,$ 
and the analogous parameter  $\,\theta _{2}\,$ of $\,SU(2)\,$. 
In a satisfactory theory these parameters
should not appear as totally free, and there should be a way to
understand the values taken.

\vskip .2truecm

This is at the starting point of the grand-unification approach~\cite{gut}, 
in particular. Can one relate the values of the three
gauge coupling constants of $\,SU(3)$, $\,SU(2)\,$ and $\,U(1)$\,? and 
establish relations between quark and lepton masses\,?
In grand-unified theories of strong, electromagnetic and weak interactions, 
$\,SU(3)$, $\,SU(2)\,$ and $\,U(1)\,$ appear as three subgroups 
of a single gauge group \,-- like $\,SU(5)\,$ or $\,O(10)$ --\,
so that the three gauge couplings of the standard model can be related.
This grand-unification of strong, electromagnetic and weak interactions 
should typically occur at very high energies initially thought to be 
of the order of $\ 10^{14}\,$ GeV\, and now, more likely,
in the framework of supersymmetric theories, $\,10^{16}\,$ GeV.
~Quarks are related with leptons, and the proton is generally expected 
to be unstable, but with a very long lifetime. 
Independently of this idea of 
grand-unification, the Higgs bosons and their interaction potential 
may be constrained in the framework of supersymmetric theories, 
in which bosons and fermions are related, and where one is led to postulate
the existence of new superpartners for all particles~\cite{alg,ssm}. 
Supersymmetric and grand-unified theories will be discussed
in section \ref{sec:symgut}.

\subsection{The problem of quantum gravity}

Gravitational interactions are well described, 
classically, by the theory of general relativity, for which the space-time 
is no longer flat but curved, the sources of the curvature being 
the densities and fluxes of energy and momentum. 
But gravity poses a severe problem when one tries to include it 
within the framework of quantum physics. Quantum gravity is not a 
renormalizable theory (nor a finite one), since the Newton constant 
$\,G_{N}\,\simeq \, 10^{-\,38}\ \hbox{GeV}^{-\,2}\,$ 
has dimension $\,-\,2$, ~in contrast with the gauge coupling constants, 
which are dimensionless.
The higher orders we go in
perturbation theory, the more and more divergent are the various successive 
amplitudes we would like to evaluate.

\vskip .2truecm

Of course for individual particles gravitational interactions are
essentially negligible at all energies up to several \,TeV's and 
even much more,
due to the extremely small value of the Newton constant. 
But since gravitational interactions act on energies and momenta,
their effective intensity behaves roughly like $\ G_{N}\,E^2\,$,
~growing quadratically with the energy. 
They should then have a strong intensity $\,(\,\sim\,1\,)\,$
at huge energies of the order of the Planck energy,  
$\ E_{P} = \sqrt{\frac{\,\hbar\ c^5\,}{G_N}}\,\simeq  \,10^{19}\ $GeV.
~The troublesome ill-defined quantum gravity effects are expected to become
important, and even essential, at such huge energies.
It is, however, very tempting to ignore them completely at much lower energies. 
Still one has to deal with this very fundamental question,
but finding a consistent quantum
theory of gravity remains an extremely hard problem. One 
hopes to deal with this difficulty by abandoning the idea 
of pointlike particles, in favor of extended objects like 
strings~\cite{strings} or membranes. 
In any case this indicates that the standard  model,
even extended to include classical gravity, 
cannot be regarded as a  satisfactory theory of all interactions. 

\vskip .2truecm

Supersymmetric theories offer a natural way to introduce gravitation 
in particle physics, in the sense that a supersymmetric 
theory of particles, once the supersymmetry is realized locally, 
necessarily includes general relativity, 
and describes also the gravitational interactions of the particles
considered. 
Furthermore supersymmetry seems to be a necessary ingredient 
for the consistency of string theories and other theories of extended objects, 
that might provide a solution to the problems of quantum gravity.

\subsection{Another problem of gravity:
the value of the cosmological constant $\,\Lambda$}

In general relativity, the energy density of the vacuum has a
definite meaning, since it couples to gravity; it can be defined and 
parametrized through a quantity called the cosmological constant, 
$\,\Lambda = \,8\,\pi\ G_N\ \rho_{\hbox{\tiny{vac}}}\,$. 
Cosmological constraints on $\,\Lambda\,$ are very severe, 
even more if they are expressed in terms of the natural
length scale associated with general relativity and gravitation, 
$L_{\hbox{\tiny{Planck}}}\ =\ \hbar/\, (m_{\hbox{\tiny{P}}}\,c)\
\simeq \, 1.6 \ \,10^{-33}\,$ cm:
\be
\vert \Lambda \vert
\ <\ \left( \,6 \ 10^{9} \,\hbox{light-years}\,\right)^{-2}\ 
\simeq \ 3 \ 10^{-56}\ {\rm cm}^{-2}
\ \simeq \ 10^{-121}\ L_{\hbox{\tiny{Planck}}}^{\ -2}\ \ .
\ee
This constraint, which corresponds to 
$\,|\,\Omega_\Lambda\,|\,=\,|\,\rho_{\hbox{\tiny{vac}}}/\rho_c\,| \,<\,2\,$,
~implies that 
\linebreak
the corresponding ``vacuum energy density'' 
$\,\rho_{\hbox{\tiny{vac}}}\,$ 
should be smaller than 
\linebreak
about twice the critical density of the Universe 
\,($\,\rho_c\,=
\,3\ H_0^{\,2}\,/\,8 \pi\,G_N\,\simeq$ $.5\ 10^{-5}\ \hbox{GeV/cm}^3\,
\simeq\,10^{-29}\ $g/cm$^3$, ~with $c=1\,$), i.e.
\be
|\,\rho_{\hbox{\tiny{vac}}}\,|\ \ <\ \ 10^{-5}\ \hbox{GeV/cm}^3\ \ \simeq 
\ \ 10^{\,-\,46}\  {\rm GeV}^4\ \ .
\ee

But before that, as the Universe expanded and cooled, 
it ought to have gone through various phase transitions. 
During  the electroweak transition,
$\,\rho_{\hbox{\tiny{vac}}}\,$ 
should have changed by some $10^{\,8}\  {\rm GeV}^4$, 
and again by some 
$10^{\,-\,4} \ {\rm GeV}^4$ during the QCD ``deconfining'' transition.
Furthermore, the na\"{\i}ve expectation would normally have been  
$\sim 10^{\,76} \ {\rm GeV}^4$, with the
vacuum quantum fluctuations of gravity cut-off at the Planck scale. Why is the
current value of the vacuum energy density so close to zero, 
compared to any of the above natural scales\,? 
Some principle seems necessary 
to explain the very small value of the cosmological constant $\,\Lambda\,$.
Here again supersymmetry may have a role to play, since bosons and fermions 
give opposite contributions to the vacuum energy density
so that the cosmological 
constant naturally vanishes in a supersymmetric theory, although the 
problem generally reappears when the supersymmetry gets broken.

\subsection{The problem of the hierarchy of mass scales}

This ``hierarchy problem'' is not a problem of the standard model 
itself, but a difficulty encountered when one tries to embed it
into a bigger theory involving large mass scales,
such as the grand-unification or the Planck scales.
These large scales tend to contaminate the electroweak 
scale associated with the $\,W^\pm $ and $\,Z\,$ bosons, 
which would make the latter very heavy. 
Indeed the parameter $\,\mu\,$ of the standard model 
which enters in the determination of the electroweak scale then tends 
to be naturally very large,
since it usually appears as an algebraic sum of extremely large quantities.
How can $\,\mu\, $ be so small with respect to much larger scales\,?
This generally requires unnatural adjustments of
parameters (first performed at the  tree approximation, 
then redone again, order by order in perturbation  theory), 
known  as ``fine-tuning". 
This is rather unsatisfactory, unless it may be
cured by means of some  principle (as in supersymmetry) allowing
scalar fields to remain light. 

\vskip .2truecm

We shall not discuss the attempts which have been made, 
as in technicolor and other models, to avoid 
the introduction of fundamental spin-0 Higgs fields 
responsible for the electroweak breaking 
(these fields being replaced by bilinear products of fermion fields 
acquiring non-vanishing  vacuum expectation values).
Technicolor models, in particular, face a number of difficulties,
and do not seem in agreement with experimental data.  
Nor shall we discuss the possibility that quarks and leptons, and 
maybe even gauge bosons, might be composite, 
since it seems difficult to pursue very far in this direction.
We shall now concentrate on the approaches of supersymmetry
and grand unification, which seem 
the most promising ones to overcome the various limitations
of the standard model, despite the fact that no direct experimental 
confirmation of these ideas has been found yet.

\vbox{\vskip.2truecm}

\section{About Supersymmetry, ~and Grand Unification}
\label{sec:symgut}

\subsection{Supersymmetry}

Supersymmetry is at first an algebraic structure~\cite{alg}, 
involving a spin-$\frac{1}{2}\,$ fermionic symmetry generator $\,Q\,$ 
satisfying the algebra:
\begin{eqnarray}
\cases{ \ \ \ba{cccc}
\{ \ Q , \ {\bar Q} \ \} \ \ &=& \
- \ 2\ \,\gamma_{\mu} \  P^{\mu} &, \vspace {0.3 true cm} \cr 
[ \,\ Q, \ P^{\mu} \ ] \ \ &=& 
  \ \ \ \ \ \ \ \ \ \ \ \ 0  \ \ \ \ \ &.
\ea
}\label{ss}           
\end{eqnarray}
This spin-$\frac{1}{2}\,$ supersymmetry generator $\,Q\,$ 
can potentially relate bosons and fermions in a physical theory, 
provided we succeed in identifying physical bosonic
and fermionic fields that might be related under such a symmetry.
The presence of the generator of spacetime translations $\,P^\mu\,$ 
in the right handside of the anticommutation relations 
is at the origin of the relation of supersymmetry with general relativity 
and gravitation, since a locally supersymmetric theory 
must necessarily be invariant under local coordinate transformations.

\vskip .2truecm
But which bosons and fermions could be related by such a supersymmetry\,? 
Perhaps the photon with the neutrino (more precisely, 
one of the three neutrinos $\,\nu_e,\ \nu_\mu\,$ and $\,\nu_\tau$,
~or conceivably a linear combination of them)\,? 
or the charged $\,W^-\,$ with the electron\,? or maybe gluons with 
\hbox{quarks ...\,?}
Alas, we quickly find out that the supersymmetry algebra is of no use to 
establish direct relations between known bosons and fermions, 
such as those we just mentioned.
If we do not want to abandon this otherwise rather appealing idea, 
we have to make a further physical assumption, 
the {\it \,superpartner \,hypothesis}~\cite{ssm}.
The photon cannot be related with any of the known neutrinos, but with a 
``neutrino'' of a new type, in some sort a ``neutrino of the photon'', 
which I called the {\it \,photino}.
In a similar way gluons have to be associated with a color-octet of 
 (self-conjugate) spin-$\frac{1}{2}$ Majorana fermions 
called the {\it \,gluinos\,} --  in spite of the old prejudice,
now forgotten, according to which no such particles should exist.
Leptons and quarks should be associated 
with new bosonic partners taken as spin-0 (rather than spin-1) 
particles, \,called {\it \,sleptons\,} and {\it \,squarks}.

\vskip .2truecm

Altogether all ordinary particles (quarks, leptons, 
gauge bosons, and Higgs bosons if they do exist) 
should be associated with new ones, still unobserved, 
that we call their superpartners.
Furthermore, in addition to this general doubling 
of the number of particle states, 
the spontaneous breaking of the electroweak symmetry 
should be induced in this framework, 
not by a single doublet of Higgs fields as in the Standard Model, 
but by a pair of them~\cite{ssm}. This implies the existence of a pair 
of charged Higgs bosons ($H^\pm$), and of several neutral ones.
Altogether this leads to what is known as the
``Supersymmetric Standard Model'', describing the 
interactions of the 8+3+1 gauge superfields of 
$\ SU(3) \times SU(2) \times U(1)$ with quark and lepton superfields, 
and two doublet Higgs superfields responsible for quark, lepton, 
and $\,W^\pm\,$ and $\,Z\,$ masses. 
Its minimal particle content is
given in Table \ref{tab:SSM}  (ignoring for simplicity 
further mixings between the various ``neutralinos''
described by mixtures of neutral gaugino and higgsino fields). 
Note that each quark $\,q\,$ or 
charged lepton $\,l\,$ of spin $\,\frac{1}{2}\,$ 
is associated with {\it \,two\,} spin-0 partners collectively denoted by
$\,\tilde q\,$ or $\,\tilde l\,$, ~while a left-handed neutrino $\,\nu_L\,$ 
is associated with a {\it \,single\,} sneutrino $\,\tilde \nu$.

\vskip 0.2cm

\begin{table}[t]
\caption{Minimal particle content of the Supersymmetric Standard Model.
\label{tab:SSM}}
\vspace{0.2cm}
\begin{center}
\begin{tabular}{|c|c|c|} \hline 
&&\\ [-0.2true cm]
Spin 1       &Spin 1/2     &Spin 0 \\ [.1 true cm]\hline 
&&\\ [-0.2true cm]
gluons ~$g$        	 &gluinos ~$\tilde{g}$        &\\
photon ~$\gamma$          &photino ~$\tilde{\gamma}$   &\\ 
------------------&$- - - - - - - - - - $&--------------------------- \\
 

$\begin{array}{c}
W^\pm\\ [.1 true cm]Z \\ 
\\ \\
\end{array}$

&$\begin{array}{c}
\ \hbox {winos } \ \widetilde W_i^\pm \\ 
[0 true cm]
\hbox {zinos } \ \widetilde Z_i \\ 
\\ 
\hbox {higgsino } \ \tilde h^0 
\end{array}$

&$\left. \begin{array}{c}
H^\pm\\
[0 true cm] H\ \\
\\
h, \ A
\end{array}\ \right\} 
\begin{array}{c} \hbox {Higgs}\\ \hbox {bosons} \end{array}$  \\ &&\\ 
[-.1true cm]
\hline &&

\\ [-0.2cm]
&leptons ~$l$       &sleptons  ~$\tilde l$ \\
&quarks ~$q$       &squarks   ~$\tilde q$\\ [-0.3 cm]&&
\\ \hline
\end{tabular}
\end{center}
\end{table}

\vskip 0.2 truecm
One of the problems one had to face, in the early days of supersymmetry, 
before it could be applied to the description of the real world, 
as indicated above,
was the fact that baryon and lepton numbers, perfectly conserved 
in all known physical processes, are carried by fundamental
fermions only (the familiar spin-$\frac{1}{2}$ quarks and leptons),
~and {\it \,not\,} by fundamental bosons. 
This feature cannot be maintained in a supersymmetric theory.
To overcome the obstacle it appeared necessary to attribute baryon 
and lepton numbers to bosonic fields as well as to fermionic ones. 
These newly-introduced bosonic fields are precisely 
the squark and slepton fields $\,\tilde q\,$ and $\,\tilde l$,  
~a denomination which makes obvious 
the fact that they must carry baryon and lepton numbers.

\vskip 0.2 truecm

Still the consideration of such new bosonic fields 
carrying baryon or lepton numbers
can be a source of additional difficulties. 
In particular known interactions are due to the exchanges of spin-1
intermediate gauge bosons (gluons, photons, $\,W^\pm$'s and $Z$'s), 
not of spin-0 particles.
In the presence of many new spin-0 particles carrying baryon and lepton 
numbers one runs into the risk of generating additional unwanted interactions 
mediated by these new \hbox{spin-0} bosons. 
This problem is naturally avoided, however, if the new squarks and sleptons 
have {\it \,no direct Yukawa couplings\,} to ordinary fermions 
(quarks and leptons)~\cite{ssm}.
This is indeed the case if the Lagrangian density involves only 
{\it \,even\,} functions of the various quark and lepton superfields,
as that must be if we are to define conserved baryon and lepton numbers.
This can be formulated in terms of a new symmetry principle, 
related with the definitions of baryon and lepton numbers.

\vskip 0.2 truecm

We were thus led to distinguish between two large classes of particles,
depending on the parity character of a new quantum number called 
$\,R\,$.
~Ordinary particles are $\,R$-even, and their superpartners $\,R$-odd
\,-- the supersymmetry generator relating the two classes of particles
being itself an $\,R$-odd operator.
The corresponding multiplicative quantum number, called 
$\,R$-parity, is  $\,+\,1\,$ for ordinary particles, $\,-\,1\,$
for their superpartners.
~This original definition can be reexpressed easily 
in terms of the spin $\,S$ of a particle, ~and of baryon and lepton numbers, 
as follows:
\be
\label{rp}
R\hbox{-parity}\ \ =\ \ (\,-\,1\,)^{2\,S\,}\ \,
(\,-\,1\,)^{\,3\,B\,+\,L}\ \ .
\ee
$R$-parity, if it actually is a symmetry of the Lagragian density, 
forbids direct couplings of the new spin-0 squarks 
and sleptons to ordinary spin-$\frac{1}{2}$ quarks and leptons. 
Expression (\ref{rp}) of $\,R$-parity illustrates that violations 
of the $\,R$-parity symmetry would necessarily imply violations 
of baryon and/or lepton number conservation laws, with the risk of 
generating unwanted processes such as 
a much too fast proton decay mediated by $\,R$-odd squark exchanges, 
if both $\,B$ and $\,L\,$-violations are simultaneously 
allowed\,\footnote{ Of course in grand-unified theories, 
in which quarks are related to leptons, 
$\,B\,$ and $\,L\,$ no longer have to be separately conserved, 
and the proton is normally expected to decay, into $\,\pi^0\,e^+$ 
~for example. But $\,B\,$ and $\,L\,$-violating processes are then mediated 
by new particles having very large masses of the order of the 
grand-unification scale, so that the resulting proton lifetime is very long.
This would not be the case if $\,B\,$ and $\,L\,$ violations 
were induced by $\,R$-odd particles with much smaller masses,
so that such violations cannot be tolerated.
Let us note, in addition, that the equivalent expression
\bea
R\hbox{-parity}\ \ =\ \ (\,-\,1\,)^{2\,S\,}\ \ (\,-\,1\,)^{\,3\,(B\,-\,L)}
\nonumber
\eea
shows that $\,R$-parity may still be conserved 
even if $\,B\,$ and $\,L\,$ are separately violated,
as long as their difference $\,B-L\,$ remains conserved, 
even only modulo 2.
\smallskip
}.

\vskip .2truecm
Whether or not it turns out to be absolutely conserved, $\,R$-parity is
essential in the discussion of the physics of supersymmetric theories.
Superpartners can only be pair-produced if $\,R$-parity 
is conserved. (Even if $\,R$-parity were not conserved 
by some of the interaction terms in the Lagrangian density, 
superpartners would still be expected to be pair-produced
in most cases.)
Furthermore, most superpartners are in general expected to be unstable, 
with ($R$-parity-conserving) decay modes such as
\be
\ba{c}
\tilde l\ \, \to\, \ l\ +\ \hbox{neutralino} \ \ ,   \vspace{2mm}\\
\tilde q\ \, \to\, \ q\ +\ \hbox{gluino}\ \hbox{(or neutralino)}\ \ ,\ \ 
\ \hbox{gluino}\ \, \to\, \ q\ \,\bar q\ +\ \hbox{neutralino}\ \ ,
\ \ \hbox{etc.,}
\ea
\ee
for example.

\vskip .2truecm

An absolute conservation of $\,R$-parity also implies 
that the lightest superpartner, called the LSP, should be absolutely stable. 
This one is presumably neutral,
the best candidate being a photino or more generally 
is a mixture of the various neutral spin-$\frac{1}{2}$ fermions 
called neutralinos.
(A stable positively charged LSP, on the other hand, would lead 
to superheavy isotopes of hydrogen, 
which have not been observed~\cite{superh}.)
This LSP is a very good candidate to constitute, at least for a part, 
the non-baryonic {\it Dark Matter\,} that seems to be present in our Universe.
{\it Missing energy\,} carried away by two unobserved LSP's is one of the most 
characteristic features of the pair production of ``supersymmetric particles'',
if $\,R$-parity is conserved.

\vskip .2truecm

One can still consider the possibility of $\,R$-parity-violating theories 
in which $\,B\,$ would be conserved but not $\,L\,$ (or conversely), 
although this would imply a somewhat dissymmetric treatment of quarks 
and leptons.
The proton would then still remain stable,
despite the presence of $\,R$-parity-violating interactions.
Let us mention, since this will be discussed in this Conference, 
that the squarks of a supersymmetric theory with $\,R$-parity violations 
have been advocated as a possible explanation  
for the so-called ``excess of high-$Q^2\,$ events'' 
in HERA experiments~\cite{hera} 
of high-energy scatterings 
of positrons and protons. Such an excess, 
if experimentally observed in a convincing way, might conceivably
be attributed to the virtual effects of 
squarks in a $\,R$-parity-violating supersymmetric 
theory, according to
\be
e^+\ \ +\ \ d^{\,-\,\frac{1}{3}}\ \ \to \ \ \hbox{virtual}\ \ 
\tilde t^{\,\frac{2}{3}} \ \
(\,\hbox{or} \ \ \tilde c^{\,\frac{2}{3}}\,) \ \ \to\ \ 
e^+\ \ +\ \ d^{\,-\,\frac{1}{3}}\ \ ,
\ee
for example.
But let us close this parenthesis, 
since the reality of the effect has not been confirmed.

\vbox{\vskip 0truecm}

Altogether a large number of experimental studies have been devoted,
since 1978, to searches for these new particles, most notably at $\,e^+e^-$ 
and $\,p\,\bar p\,$ colliders.
The quest still remains unsuccessful, which implies lower limits 
on superpartner masses, generally of the order of $\,70-80$ GeV$/c^2\,$ 
for the various sleptons and winos, and more than a hundred \,GeV$/c^2\,$'s, 
~for the strongly-interacting squarks and gluinos~\cite{pdg}.
When can we finally expect to detect these long-awaited superpartners\,?  
We may still have to wait for the Large Hadronic Collider at CERN, 
or maybe even a future very-high-energy linear $e^+ e^-$ collider.
But the mass scale at which these new particles 
should be found is normally expected to be of the order of the 
electroweak scale, or up to a few TeV/$c^2\,$ at most, if we do not want a
new large mass scale associated with the superpartners to {\it \,create\,} 
a hierarchy problem in the electroweak theory.

\vskip 0.5 true cm

\subsection{Grand unification}

Grand-unification theories~\cite{gut} are invariant under larger gauge groups
such as the $\,SU(5)\,$ of Georgi and Glashow
(considered as the prototype of a grand-unification 
gauge group), or bigger groups like $\,O(10)\,$ or $\,E(6)\,$. 
They contain as subgroups both the color $\,SU(3)\,$
gauge group of the strong interactions and the $\,SU(2)\times U(1)\,$
gauge symmetry group of electroweak interactions.
This leads to the possibility of relating the three corresponding gauge 
couplings, $\,g_s$, $\,g\,$ and $\,g'\,$.

\vskip .2truecm

Once the various symmetry generators are normalized 
according to the same convention,  
which leads to consider the suitably-normalized gauge couplings
\be
\label{ggg}
g_3\ \equiv \ g_s\ ,\ \ \ \ g_2\ \equiv\ g\ \ \ \ \hbox{and}\ \ \ \
g_1\ =\ \sqrt{\,\frac{5}{3}}\ \ g'\ \ ,
\ee
the grand-unification symmetry imposes, above an appropriate energy scale 
called the grand-unification scale ($\,m_X\,$), 
~the equality of the three couplings $g_{i=1,2,3}\,$.
This would imply, in particular, 
$\ \tan\,\theta\,=\,g'/g\,=\,\sqrt{\,3/5\,}\,$, 
and therefore\,\footnote{ This value of $\,\sin^2\theta\,$  
may be obtained directly by expressing that 
$\,g\,T_3\,$ and $\,e\,Q\,$ appear as two $SU(5)\,$ generators 
with the same normalisation, 
which requires that $\,g^2\ \hbox{Tr}\ T_3^{\ 2}\,=\,e^2\ \hbox{Tr}\ Q^2\,$.
Computing the trace for the fifteen chiral quark and lepton fields 
in one family, we have $\ \hbox{Tr}\ T_3^{\ 2} = 8\ \frac{1}{4}=  2\,$, 
$\,\hbox{Tr}\ Q^2 = \frac{16}{3}\,$,
~and therefore $\,\sin^2\theta\,=\,e^2/g^2\,=\,3/8\,$, 
~at the grand unification 
scale.}
\be
\sin^2\theta\ \ =\ \ \frac{3}{8}\ \ . 
\ee

But the equality of the three gauge couplings (\ref{ggg}) 
\,-- and the resulting value $\,\sin^2\theta\,=\,3/8$ --\, 
should only be valid 
at the very high grand-unification scale, at which electroweak 
and strong interactions are expected to have 
the same intensity. This is of course not the case at present energies, 
but one has to take into account the evolution of the three 
gauge couplings between their ``low-energy'' values, 
presently known at the energy scale
of about 100 GeV, and the very high-energy scale $\,m_X\,$
at which the grand unification would start to become manifest.

\vskip .2truecm

All three gauge couplings are in  fact slowly-varying functions of the energy, 
whose values may be extrapolated from the present ``low'' energies 
(i.e. $\approx\,m_Z$) to very high energies, if the particle spectrum is known. 
For any given particle content assumed to be known one can derive,
from the values of $\ \alpha(m_Z)\,$ and $\,\alpha_s(m_Z)$,
~the expected values of the electroweak 
mixing angle $\,\theta\,$ and of the grand-unification scale $\,m_X$, 
~an essential ingredient in the determination of the expected lifetime 
of the proton.

\vskip .2truecm

The minimal version of the $\,SU(5)\,$ model came close to a successful 
determination of $\,\sin^2\theta\,$ ($\,\simeq .214 \pm  .004$), 
but there is now a significant discrepancy with the measured value,
$\,\sin^2\theta \simeq .231\,$. In other terms the three gauge couplings, 
determined from their now well-known ``low-energy'' values
(with $\,\sin^2 \theta\,(m_Z) \simeq .231\,$,
$\,\alpha(m_Z)\,\simeq\,1/129,\ \,\alpha_s(m_Z)\,\simeq\,.12\,$), 
fail to meet at a common unification scale: 
there is no high-energy unification of the couplings. 
Furthermore the partial lifetime for the decay of the proton into 
$\,\pi^0\,e^+\,$ was initially expected around 4 $10^{29}$ years,
in the simplest $\,SU(5)\,$ model
(for which  $\,m_X\,\simeq \,(\,1\,$ to $\,4\,)\ 10^{14} \,$ GeV$/c^2\,$), 
with an estimated upper bound of about \hbox{3 $10^{31}$} years. 
But experiments found no candidate for this decay mode, 
which implies a partial lifetime larger than $\,10^{33}\,$
 years~\cite{pdecay}.
This would require the grand-unification mass $\,m_X\,$ 
to be larger than $\,\simeq \,10^{15}\,$ GeV$/c^2$, a result
also in disagreement with the minimal version of the $\,SU(5)\,$ model.

\vskip .2truecm

The situation changes drastically, however, 
if the effects of the new particles of the supersymmetric standard model 
are taken into account.
They lead to modified expectations for the values 
of the electroweak mixing angle $\,\theta\,$ 
and of the grand-unification mass $\,m_X\,$~\cite{unif}.
The unification scale is increased to about $\ \approx 10^{16}\,$ GeV$/c^2$,
~mostly as an effect of the gluinos and the $\,SU(2)\,$ gauginos, 
which slow down or even reverse the evolution of the non-abelian gauge 
couplings $\,g_3\,$ and $\,g\,$ of $\,SU(3)\,$ and $\,SU(2)$. 
At the same time the two Higgs doublets, with their associated higgsinos, 
modify the expected value of  $\ \sin^2\theta$, 
~making it in good agreement with the one experimentally measured.

\vskip .2truecm

     In the framework of grand-unification we now have an indirect 
indication that we may be on the right track. Superpartners 
and Higgs bosons would have a crucial influence on the evolution 
of the three $\,SU(3) \times SU(2) \times U(1)$ gauge couplings, and are 
essential to obtain their convergence at high energy -- around $10^{16}$ 
GeV or so -- with the measured values of $\alpha_s(m_Z) \simeq 0.12$ and
${\sin}^2 \theta \simeq 0.231\,$. This convergence, obtained
for the particle content of the Supersymmetric  Standard Model with 
the new particles at their expected mass scale of 
$\ \,\approx 100$ GeV\, to \,1 TeV\, or so, may be taken, 
optimistically, as an indication in favor of the existence of these 
superpartners, and as a unification test of the Supersymmetric Standard 
Model.

\vbox{\vskip -.2truecm}

\section{About new neutral currents, and (very) light gauge bosons}
\label{sec:newbos}

\vskip .1truecm

\subsection{Searching for new neutral gauge bosons}
\label{subsec:lightgauge}

A large amount of very precise data has been accumulated over the years from 
parity-violation atomic physics experiments, 
polarized electron-nucleon scattering experiments, 
neutrino-scattering and
$\,e^+ e^-$ and $\,p\,\bar p\,$ scattering experiments, etc..
These data allow for very precise tests of the standard model, 
and a precise determination of $\,\sin^2 \theta\,$ which is
essential for the discussion of grand-unified theories, as we just saw.
They also lead to restrictions on the existence of 
possible new neutral gauge bosons $\,Z'\,$ 
coupled to additional weak neutral currents.

\vskip .2truecm

One is generally used to think of new neutral gauge bosons as being very heavy,
as they should in left-right symmetric models (with
gauge group $\,SU(2)_L \times  SU(2)_R \times U(1)_{B-L}\,$), 
or in grand-unified theories with larger gauge groups 
such as $\,O(10)\,$ or $\,E(6)$.
All such models include at least one additional neutral gauge boson $\,Z'$, 
with gauge couplings to ordinary fermions
roughly of the order of the $\,SU(2) \times U(1)\,$ gauge couplings
$\,g\,$ and $\,g'$, ~as for the $\,Z\,$ couplings.
There are, also, additional charged $\,W'^\pm\,$ bosons 
coupled to right-handed currents.
From the non-observation of such new $\,Z'\,$ bosons 
in $\,p\,\bar p\,$ scattering experiments at Fermilab, one can deduce 
lower limits on their masses, which may reach about \,600 GeV$/c^2$,
depending on the hypothesis made. Similar limits also exist
 for the additional $\,W'^\pm$'s.

\vskip .2truecm
It is worth to keep in mind, however, 
the much less conventional possibility 
of new neutral gauge bosons that we call $\,U\,$ bosons, 
that could be much lighter, and possibly even extremely light, 
provided their gauge coupling $\,g"\,$ is, at the same time, 
relatively weak, or even extremely weak~\cite{U}. 
In the case of light $\,U $ bosons we can in general no longer evaluate 
exchange amplitudes in  the local limit approximation
as we often do for $\,Z\,$ bosons,
and it is essential to take into account propagor effects.

\vskip .2truecm

To compare the magnitudes of $\,Z$-exchange and $\,U$-exchange amplitudes, 
let us recall that neutral current effects of the $\,Z\,$ boson 
are fixed 
by
\be
\frac{g^2\ +\ g'^2}{q^2\,-\,m_Z^{\ 2}}\ \ \simeq \ \ 
-\ \frac{8\ \,G_F}{\sqrt 2}
\ \ \simeq\ \ -\ \frac{4}{v^2} \ \ ,
\ee
which is proportional to $\,G_F$,
~in the local limit approximation for which $\ |\,q^2|\, \ll\,m_Z^{\ 2}\,$.
$\ v\,=\,(G_F\,\sqrt 2)^{-\frac{1}{2}}\simeq\, $ 246 GeV
is often called the electroweak symmetry-breaking scale (c.f. footnote
\,{\em \ref{pothiggs}}\  ~in section \ref{sec:electroweak}).

\vskip .2truecm

In a similar way, for a {\it \,heavy\,} $\,U\,$ boson 
with gauge coupling $\ g"$,
~and a mass $\,m_U\,$ written proportionally to $\,g"\,F$, $\,F\,$ denoting
the extra-$U(1)\,$ symmetry-breaking scale (say $\ m_U =\,g"\,F/2\,$,
~to keep things simple),
~the corresponding amplitude may be written similarly, 
again in the local limit approximation, proportionally to:
\be
\frac{g"^2}{q^2\,-\,m_U^{\ 2}}\ \ \simeq \ \ -\ \,\frac{4}{F^2}\ \ \ \ 
(\hbox{at lower}\ \ |\,q^2| \,\ll \,m_U^{\ 2}\,)\ \ .
\ee
$U$-exchange amplitudes are then of the same order of magnitude 
as $\,Z$-exchange amplitudes, up to a scale factor 
\be
r^2\ \ =\ \ \frac{v^2}{F^2}\ \ =\ \ \left(\ \frac{\hbox{electroweak scale}}
{\hbox{extra $\,U(1)\,$ symmetry breaking scale}}\ \right)^2 \ \ .
\ee
This ratio may be rather small, provided the corresponding new symmetry 
is broken at a scale $\,F\,$ sufficiently high 
compared to the electroweak scale $\,v\ \simeq\,250\,$ GeV, 
as it happens if an extra Higgs singlet 
acquires a large vacuum expectation value $\ \approx\,F\, \gg \,v\,$.
(When the new coupling $\,g"\,$ 
is of the same order as $\,g\,$ or $\,g'$, ~this simply means,
not surprisingly, that 
$\,U$-exchange amplitudes are small compared to $\,Z$-exchange amplitudes, 
provided the $\,U\,$ boson is sufficiently heavy compared 
to the $\,Z\,$.)

\vskip .4truecm

The situation is, however, different in the case of relatively light 
or very light weakly-coupled $\,U\,$ bosons, 
for which the local limit approximation 
is in general no longer valid~\cite{U}. We then have (in magnitude)
\be
\label{prop}
\frac{g"^2}{q^2\,-\,m_U^{\ 2}}\ \ =\ \ \frac{g"^2}{m_U^{\ 2}}\ \ 
\frac{m_U^{\ 2}}{q^2\,-\,m_U^{\ 2}}\ \ 
\simeq\ \ \frac{g"^2}{q^2}\ \ 
\ll \ \ \frac{4}{F^2}\ \ \ \ \ \hbox{for}\ \ m_U^{\ 2}\,\ll\, |\,q^2|\ \ .
\ee
As a result 
experiments performed at higher values of $\,|\,q^2|\,$ 
(compared to $\,m_U^{\ 2}\,$)
may well be insensitive to the existence of such light $\,U\,$ bosons,
while experiments performed at lower $\,|\,q^2|\, \simle m_U^{\ 2}$,
~such as for example atomic physics experiments, 
would have a much better sensitivity.
As an illustrative example for an extra $\,U(1)\,$ symmetry broken 
at or around the electroweak scale (i.e. $F\,\approx \,v$), 
~atomic physics experiments 
performed with heavy atoms can be quite sensitive to the parity-violation 
effects induced by a $\,U\,$ boson, provided its mass is larger than about 
\,1 MeV$/c^2\,$~\cite{pvat}. ~Such a $\,U\,$ boson, on the other hand,
could have escaped detection in scattering experiments 
performed at higher values of $\,|\,q^2|$,
owing to the $\,m_U^{\ 2}/(q^2 - \,m_U^{\ 2})\,$ reduction factor 
in the expression of $\,U$-exchange amplitudes.

\vskip .2truecm

The situation, however, again deserves further attention 
if the new gauge boson is 
really a {\it very\,} light $\,U\,$ boson, so that $\,m_U^{\ 2}\,$ 
is practically always negligible,
even in atomic physics parity-violation experiments. 
The amplitudes, proportional to $\,g"^2/q^2$, 
~are, again, extremely small owing to the very small value of the 
gauge coupling $\,g"$, ~and even seem to vanish in the limit $\,g"\,\to\,0\,$. 
$\,U$-boson effects would then seem to be 
totally negligible, in this limit.
This, however, is not necessarily always the case.
$\,U\,$-boson exchanges could still lead to a new long range force, 
that might be detected through apparent violations of the Equivalence 
Principle.
Furthermore particle physics experiments themselves could be sensitive 
to such particles, even if the corresponding gauge coupling 
$\,g"\,$ becomes {\it \,arbitrarily small}, \,a somewhat surprising statement\,!

\subsection{A new long-range force\,?}
 
The possible extra $\,U(1)$ symmetries that could be gauged,
in addition to the weak hypercharge $\,U(1)$, 
~depend on the set of Higgs doublets 
responsible for the electroweak breaking. 
After mixing effects with the $\,Z\,$ boson are taken into account,
the resulting $\,U\,$ current 
involves in general both {\it \,vector\,} and {\it \,axial\,} parts.
The vector part is generally associated with the charge 
\begin{equation}
Q_5 \ =\  x \,B \,+ \,y_i \,L_i \,+ \,z\, Q_{el}        \ \ .
\end{equation}
This vector part in the $\,U\,$ current would be responsible for a new force, 
acting on ordinary neutral matter in an {\it additive\,} way, 
proportionally to a linear combination of the numbers of protons 
and neutrons, $\,Z\,$ and $\,N$. 
~If the $\,U$ boson is massless or almost massless with
an extra $\,U(1)\,$ gauge coupling $\,g"$ extremely small, 
the new force would superpose its effects to those of gravitation, 
leading to apparent violations of the Equivalence Principle,
since the numbers of neutrons and protons in an object 
are not exactly proportional to its mass~\cite{U,newforce}.
Newton's $\,1/r^2\,$ law of gravitation could also seem to be
violated, if the new force has a finite range.

\vskip .2truecm

If they have an appropriate magnitude, 
such violations of the Equivalence Principle
could be detectable by the STEP experiment~\cite{step} 
(Satellite Test of the Equivalence Principle).
Violations of this principle could also be due 
to massless or quasimassless spin-0 particles, such as the ``dilaton'' of 
some superstring inspired models~\cite{damour}.
By monitoring the relative motion of two test masses
of different compositions\,\footnote{ The test masses, however, 
cannot be taken as spherical, but only cylindrical.
A potential difficulty is the existence of residual interactions 
between their higher multipole moments 
and the gravity gradients induced by disturbing masses within the satellite. 
This could simulate a ``violation of the Equivalence Principle''.
To minimize these effects one can use test masses approaching ideal forms of 
``aspherical gravitational monopoles'',
which are (homogeneous) solid bodies for which {\it \,all\, higher 
multipole moments vanish identically,\,} despite the lack of spherical 
symmetry~\cite{cdf}.
} 
circling around the Earth, in a drag-free satellite,
this experiment aims at testing the validity 
of this principle at a level of precision 
that could reach $\,\sim 10^{-\,17}\,- \,10^{-\,18}$,
~an improvement of five orders of magnitude compared to the present 
situation, in which this principle is known to be valid at a level 
of precision of about $\,10^{-\,12}$.
Testing, to a very high degree of precision, the Equivalence Principle 
in space would bring new constraints on the possible existence 
of such forces, and might conceivably lead to a spectacular discovery, 
should a deviation from this Principle be found.

\subsection{A very light spin-1 
$\ {\boldmath U}\,$ boson does ${\boldmath not\,}$ decouple for vanishing 
gauge coupling \,!}

Let us now discuss whether it could make sense to search for a spin-1
$\,U\,$ boson with an {\it \,extremely small\,} gauge coupling $\,g"$, 
~in particle physics experiments.
No one, however,
would imagine being able to search directly for {\it gravitons\,} 
in a particle physics experiment, due to the extremely small value of 
the Newton constant
($\,\simeq \,10^{-38}$, ~in units of \,~GeV$^{-2}$). 
Then how could we search directly,
in particle decay experiments, for $\,U$-bosons
with even smaller values of the corresponding coupling, 
$\ g"^2\ll 10^{-38}\ $? Still this turns out to be possible\,!
This rather astonishing result 
involves an ``equivalence theorem'' 
between the interactions of spin-1 particles and those of spin-0 particles, 
in the limit of  very small gauge couplings~\cite{U}.

\vskip .2truecm

     One might think that, in the limit of vanishing extra-$U(1)\,$ 
gauge coupling constant $\,g"$, ~the effects of the  new  gauge  boson 
would be arbitrarily small, and may therefore be disregarded. 
But in general this is {\it wrong\,}, 
as soon as the $\,U$-current involves a (non-conserved) {\it axial\,} part\,!
The amplitudes for emitting a very light ultrarelativistic $\,U\,$
boson, proportional to $\,g"$, ~seem indeed to vanish with $\,g"$. 
This is, however, misleading, since the polarization vector for
a longitudinal $\,U\,$ boson, $\,\epsilon^{\mu} \simeq \,k^{\mu}/m_U\,$, 
becomes singular in this limit: $\,m_U = \frac{1}{2}\ g"\, F\,$  also vanishes 
with $\,g"$. ~Altogether the amplitudes for emitting, absorbing, 
or exchanging a longitudinal $\,U\,$ boson become independent of $\,g"$ 
when the $\,U\,$ boson is ultrarelativistic. Such a $\,U\,$ boson does 
then behave like a spin-0 particle\,\footnote{ The effects of the 
longitudinal ($\,q^\mu q^\nu\,$) term in the $\,U\,$ boson propagator 
are essential in the limit considered, for which $\ m_U^{\ 2}\,\ll\,|\,q^2|\,$.
$\,U$-exchange amplitudes are then proportional to
\bea
\frac{g"^2}{q^2\,-\,m_U^{\ 2}}\ \ 
(\ g^{\mu\nu}\,-\,\frac{q^\mu q^\nu}{m_U^{\ 2}}\ )
\ \ \to \ \ -\ \frac{4}{F^2}\ \ \frac{1}{q^2} \ \ q^\mu q^\nu\ \ .
\nonumber 
\eea
One reconstructs the amplitudes for exchanging a massless spin-0 Goldstone 
particle, with derivative couplings proportional to $\,1/F$.
Applied to (non-conserved) axial currents
$\ \bar f\,\gamma^\mu\,\gamma_5\,f$, ~the $\,q^\mu q^\nu\,$ terms regenerate 
the pseudoscalar Yukawa couplings of the equivalent spin-0 particle, 
proportional to $\,1/F\,$ times quark or lepton masses. \\
}.   
This ``equivalence theorem'' expresses that in the 
low-mass or high-energy limit (i.e., for $\,m_U \ll E\,$), the third degree of 
freedom of a massive $U$-boson continues to behave like the massless spin-0 
Goldstone boson which was ``eaten away''. For very small $\,g"$ 
 the  \hbox{spin-1} $\,U$-boson simply behaves as this massless spin-0 
Goldstone boson.

\vskip .4truecm

Incidentally  the same phenomenon, in the case of local supersymmetry,
called supergravity, expresses that a {\it very light\,} spin-$\frac{3}{2}$ 
{\it \,gravitino} (the superpartner of the spin-2 graviton), 
having interactions fixed by the gravitational ``gauge'' coupling constant
$\,\kappa = \sqrt{\,8\,\pi\ G_N}\,\simeq\, 4.1\ \,10^{- 19}\,$ (GeV)$^{-1}$,
~would behave very much like a massless spin-$\frac{1}{2}\,$ 
{\it goldstino}~\cite{ssm}.
~Just like the mass of the $\,U\,$ boson was given in terms of the 
extra $\,U(1)\,$ gauge coupling $\,g"\,$ and symmetry breaking scale $\,F\,$ 
by the formula
$\ m_U\,=\,\frac{1}{2}\ \,g"\,F$,
~the mass of the spin-$\frac{3}{2}$ gravitino is fixed by its 
(known) ``gauge'' coupling constant $\,\kappa\,$ and 
the (unknown) supersymmetry-breaking scale parameter $\,d$, ~as follows:
\be
m_{\hbox{\tiny{3/2}}}\ \ =\ \ \frac{\kappa\ d}{\sqrt 6}\ \ \simeq \ \ 1.68\ 
\left(\ \frac{\sqrt d}{100\ \hbox{GeV}}\right)^2\ 10^{-6}\ \hbox{eV}/c^2\ \ .
\ee

\vskip .1truecm

\noindent
The interactions of a light gravitino are in fact determined by the ratio
$\kappa/m_{\hbox{\tiny{3/2}}}\,$, ~or
$\ G_N/ m_{\hbox{\tiny{3/2}}}^{\ 2}\,$, ~so that a sufficiently light gravitino 
might be detectable in particle physics experiments, 
despite the extremely small value of the Newton constant $\,G_N\,\simeq\,
10^{- 38}\,$ (GeV)$^{-2}$,
~provided the supersymmetry-breaking scale\,\footnote{An equivalent notation 
makes use of a parameter $\,\sqrt F\,=\,\sqrt d\,/2^{\hbox{\tiny{1/4}}}$,
~defined so that $\,F^2 =\,d^2/2$, ~and therefore
\bea
m_{\hbox{\tiny{3/2}}}\ \ =\ \ \frac{\kappa\ F}{\sqrt 3}\ \ \simeq\ \ 2.37\ 
\left(\,\sqrt F \,/\, 100\ \hbox{GeV}\,\right)^2\ 10^{-6}\ \hbox{eV}/c^2 \ \ .
\nonumber 
\eea
Furthermore, the supersymmetry-breaking scale ($\,\sqrt d\,$ or $\,\sqrt F\,$) 
associated with a (stable or quasistable) light gravitino should in principle 
be smaller than a few $\,10^6$ GeV\,'s, 
~for its mass to be sufficiently small
($\,m_{\hbox{\tiny{3/2}}}\,\simle \,1 $ keV$/c^2$), so that relic gravitinos 
do not contribute too much to the energy density of the Universe.
} $\ \sqrt d\ $ is not too large.
The gravitino would then be the lightest supersymmetric particle,
with all other superpartners expected to ultimately produce a gravitino 
among their decay products, if $\,R$-parity is conserved.
(In particular the lightest neutralino could decay into photon + gravitino, 
~so that the pair-production of ``supersymmetric particles'' 
could lead to final states including two photons with missing energy
carried away by unobserved gravitinos.)

\vskip .2truecm

For a sufficiently light gravitino one can also search 
for the {\it direct production\,} of a
single gravitino associated with an unstable photino $\,\tilde \gamma$
~(or more generally a neutralino), decaying into
gravitino + $\,\gamma$, ~in $\,e^+e^-\,$ annihilations.
Or for the radiative pair-production of two gravitinos
in $\,e^+e^-\,$ or $\,p\,\bar p\,$ annihilations 
at high energies~\cite{gravitino},
e.g.
\be
e^+ e^- \ \ (\hbox{or}\ \ p\,\bar p)\ \ \ \rightarrow\ \ \ 
\gamma\ \ (\,\hbox{or jet\,)}\ \ +\ \ 
2\ \,\hbox {unobserved gravitinos}\ \ ,
\ee
which have cross-sections 
\be
\sigma \ \  \propto\ \ 
\frac{G_N^{\ 2} \ \ \alpha\, (\,\hbox{or} \ \,\alpha_s\,)\ \,s^3}
{m_{\hbox{\tiny{3/2}}}^{\ \ 4}}\ \ \propto\ \ 
\frac{\alpha\ (\,\hbox{or} \ \,\alpha_s\,)\ \,s^3}{d^4}\ \ \ .
\ee
Although the existence of so light gravitinos may appear 
as relatively unlikely, such experiments are sensitive to gravitinos of 
mass $\,m_{\hbox{\tiny{3/2}}} \simle\,10^{-5}\,$ eV/$c^2$, 
~corresponding to supersymmetry-breaking scales 
smaller than a few hundreds of GeV's.

\vskip .4truecm

Just in the same way as the magnitude of $\,\sqrt d\,$ 
determines the effective strength of the gravitino interactions
(and whether or not we may have a chance to produce it directly 
in particle physics experiments),
the magnitude of the extra $\,U(1)\,$ symmetry-breaking scale
parameter $\,F\,$ determines the effective strength of the 
$\,U\,$ boson interactions, for a $\,U\,$ coupled 
to a non-conserved current, as it is the case if this current includes 
an axial part.
\linebreak
 (A $\,U\,$ boson 
coupled to a conserved current does effectively decouple in the limit 
of vanishing gauge coupling.) \hbox{\hskip 3truecm}

\vskip .2truecm

For such an extra $\,U(1)\,$ gauge symmetry broken at the electroweak scale 
\linebreak ($\,F \approx 250\ \, GeV\,$) by two Higgs doublets,
the spin-1 $\,U$-boson acquires essentially the same 
effective pseudoscalar couplings to quarks and leptons 
as a ``standard'' spin-0 axion. 
Its existence is then excluded by the results of 
$\ \psi\ (c\bar c)\,\to \,\gamma\,+\,\hbox{``nothing''}\,$ 
and $\ \Upsilon\ (b\bar b)\,\to \,\gamma\,+\,\hbox{``nothing''}$ 
decay experiments~\cite{U,newforce,upsilon}. They imply 
that the extra-$U(1)$ symmetry should be broken at a scale $\,F$ 
at least of the order of twice the electroweak scale $\,v$.
~If, on the other hand, the  extra $\,U(1)\,$ is broken 
``at a large scale'' $\,F\,$ significantly higher than  the  electroweak
scale ($\,F\, \gg\, 250 \ GeV\,$) -- using for example a very large 
Higgs-singlet vacuum-expectation-value, 
$\,U$-boson effects in particle physics would be practically ``invisible''.
This mechanism relying on a large Higgs singlet v.e.v. and applied to 
an extra global $\,U(1)_{\rm PQ}\,$ symmetry broken at a very high scale 
also allowed us to indicate that axion interactions 
in particle physics could be made very small, 
with this axion mostly an electroweak singlet~\cite{U,invax}.

\vskip .2truecm

Such light $\,U\,$ bosons were initially
discussed in the framework of supersymmetric theories with an extra 
$\,U(1)\,$ gauge group, but their possible existence should be 
considered independently of this original motivation.
To close this special section dealing with new light bosons,
let us also mention that the exchanges of a new spin-1 $\,U\,$ boson, 
or of a spin-0 particle such as the axion, 
could lead to new forces acting on particle spins,
including a ($CP$-conserving) spin-spin interaction, 
and possibly a very small $(CP$-violating) ``mass-spin coupling'' 
interaction.

\section{Conclusion}

Experiments on Parity Violation in electron-hadron electroweak interactions 
brought spectacular contributions to our knowledge of the properties 
of the fundamental interactions, and contributed greatly to establish 
the validity of the Standard Model. 
But we certainly expect new physics beyond the Standard Model.
Very precise tests of this model give precious informations 
about the mechanism of electroweak symmetry breaking, and 
constraints on new physics beyond the standard model: 
existence of new gauge bosons, grand-unification, supersymmetry, ...
This requires the study of many different physical phenomena, 
in various domains of physics, from very low to very high energies,
and from microscopic physics to astrophysics and cosmology.
We are eagerly waiting for new experimental results, and, certainly, 
a very interesting Conference.


\section*{References}

\begin{thebibliography}{99}


\bibitem{sm1} S.L. Glashow, {\em Nucl. Phys.} {\bf 22}, 579 (1961);
A. Salam and J.C. Ward, {\em Phys. Lett.} {\bf 13}, 168 (1964);
S. Weinberg, {\em Phys. Rev. Lett.} {\bf 19}, 1264 (1967);
A. Salam, Proc. 8th Nobel Symp., ed. N. Svartholm, 367 (1968).


\bibitem{sm2} 
S. Weinberg,  {\em Rev. Mod. Phys.} {\bf 52}, 515 (1980);
A. Salam,     {\em Rev. Mod. Phys.} {\bf 52}, 525 (1980);
S.L. Glashow, {\em Rev. Mod. Phys.} {\bf 52}, 539 (1980).


\bibitem{higgs} F. Englert and R. Brout, 
{\em Phys. Rev. Lett.} {\bf 13}, 321 (1964);
P.W. Higgs, {\em Phys. Lett.} {\bf 12}, 132 (1964);
{\em Phys. Rev. Lett.} {\bf 13}, 508 (1964).


\bibitem{cn} F.J.  Hasert et al., {\em Phys. Lett.} B {\bf 46}, 121 (1973); 
B {\bf 46}, 138 (1973); 
A. Benvenuti et al., {\em Phys. Rev. Lett.} {\bf 32}, 800 (1974).


\bibitem{z} G. Arnison et al., {\em Phys. Lett.} B {\bf 126}, 398 (1983);
P. Bagnaia et al., {\em Phys. Lett.} B {\bf 129}, 130 (1983).


\bibitem{zel} 
Ya. B. Zel'dovich, {\em J. Exptl. Theoret. Phys.} {\bf 36}, 964 (1959)
[{\em Sov. Phys. JETP} {\bf 9}, 682 (1959)]; F. Curtis Michel, 
{\em Phys. Rev} B {\bf 138}, 408 (1965).


\bibitem{slac} C.Y. Prescott et al., {\em Phys. Lett.} B {\bf 77}, 347 (1978); 
 B {\bf 84}, 524 (1979).


\bibitem{bou} M.A. Bouchiat and C. Bouchiat, 
{\em Phys. Lett.} B {\bf 48}, 111 (1974); 
contribution to these Proceedings, and references therein.


\bibitem{keo} See in particular R. Mc Keown, these Proceedings, 
and references therein.


\bibitem{mesqw} C.S. Wood et al., {\em Science} {\bf 275}, 1759 (1997);
V.A. Dzuba, V.V. Flambaum and O.P. Sushkov, preprint hep-ph/9709251.


\bibitem{lan} J. Erler and P. Langacker, 
{\em Eur. Phys. Jour.} C {\bf 3}, 90 (1998).


\bibitem{musolf} M. Ramsey-Musolf, these Proceedings, and references therein.


\bibitem{altarelli} G. Altarelli, these Proceedings, and references therein.


\bibitem{nu} The Super-Kamiokande collaboration, 
{\em Phys. Rev. Lett.} {\bf 81}, 1562 (1998).


\bibitem{axion} S. Weinberg, {\em Phys. Rev. Lett.} {\bf 40}, 223 (1978);
F. Wilczek, {\em Phys. Rev. Lett.} {\bf 40},  279 (1978).


\bibitem{sak} A. Sakharov, {\em Pis'ma Zh. Eksp. Teor. Fiz.} {\bf 5}, 32 (1967)
[{\em JETP Lett.} {\bf 5}, 24 (1967)].


\bibitem{gut} 
H. Georgi and S.L. Glashow, {\em Phys. Rev. Lett.} {\bf 32}, 438 (1974);
H. Georgi, H.R. Quinn and S. Weinberg, {\em Phys. Rev. Lett.} {\bf 33},  451
(1974);
P. Langacker, {\em Phys. Reports} {\bf 72}, 185 (1981).


\bibitem{alg} Yu. A. Gol'fand and E.P. Likhtman, 
\Journal{\em ZhETF Pis. Red.}{13}{452}{1971} 
[\Journal{\em JETP Lett.}{ 13}{323}{1971}];
D.V. Volkov and V.P. Akulov, {\em Phys. Lett.} B {\bf 46}, 109 (1973);
J. Wess and B. Zumino, {\em Nucl. Phys.} B {\bf 70}, 39 (1974);
{\em Supersymmetry and Supergravity}, A Reprint Volume of Physics Reports, 
ed. M. Jacob (North Holland/World Scientific, 1986).

\bibitem{ssm} P. Fayet, {\em Phys. Lett.} B {\bf 69}, 489 (1977); 
B {\bf 70}, 461 (1977); 
in {\em History of original ideas 
and basic discoveries in Particle Physics}, 
eds. H. Newman and T. Ypsilantis,
Proc. Erice Conf., {\em NATO Series} B {\bf 352},  639 (Plenum, NY, 1996);
G.R. Farrar and P. Fayet, {\em Phys. Lett.} B {\bf 76}, 575 (1978).


\bibitem{strings} See e.g. M. Green, J. Schwarz and E. Witten,
{\em Superstring theory}, Cambridge University Press (1987).


\bibitem{superh} 
T.K. Hemmick et al. {\em Phys. Rev.} D {\bf 41}, 2074 (1990);
P. Verkerk et al., {\em Phys. Rev. Lett.} {\bf 68}, 1116 (1992).

\bibitem{hera}
H1 Collaboration, {\em Z. Phys.} C {\bf 4}, 191 (1997);
ZEUS Collaboration, {\em Z. Phys.} C {\bf 4}, 207 (1997).


\bibitem{pdg} {\em Review of Particle Properties}, 
{\em Eur. Phys. Jour.} C {\bf 3}, 1 (1998).


\bibitem{pdecay}
R. Becker-Szendy et al., {\em Phys. Rev.} D {\bf 42},  2974 (1990); 
K.S. Hirata et al., {\em Phys. Lett.} B {\bf 220},  308 (1989);
the Super-Kamiokande collaboration, {\em Phys. Rev. Lett.} {\bf 81}, 3319
(1998).


\bibitem{unif} 
See e.g. U. Amaldi, W. de Boer and H. F\"urstenau, {\em Phys. Lett.}  B
{\bf 260}, 447 (1991); 
P. Langacker and M. Luo, {\em Phys. Rev.} D {\bf 44}, 817 (1991).


\bibitem{U} P. Fayet, {\em Phys. Lett.} B {\bf 95}, 285 (1980);
{\em Nucl. Phys.}  B {\bf 187}, 184 (1981).


\bibitem{pvat} P. Fayet, {\em Phys. Lett.} B {\bf 96}, 83 (1980);
C. Bouchiat and C.A. Piketty, {\em Phys. Lett.} B {\bf 128}, 73 (1983).


\bibitem{newforce} P. Fayet, {\em Nucl. Phys.} B {\bf 347}, 743 (1990);
{\em Class. Quantum Grav.} {\bf 13}, A19 (1996).


\bibitem{damour} T. Damour and A.M. Polyakov, 
{\em Nucl. Phys.} B {\bf 423}, 532. (1994).


\bibitem{step} {\em MiniSTEP}, NASA-ESA report (1996); see also
{\em \,STEP, Satellite Test of the Equivalence Principle},
Report on the Phase A Study, {\em ESA SCI} (96) 5 (1996).


\bibitem{cdf} A. Connes, T. Damour and P. Fayet, 
{\em Nucl. Phys.} B {\bf 490}, 391 (1997).


\bibitem{gravitino} P. Fayet, {\em Phys. Lett.} B {\bf 117}, 460 (1982);
B {\bf  175}, 471 (1986); A. Brignole et al., 
{\em Nucl. Phys.} B {\bf 516}, 13 (1998); B {\bf 526}, 136 (1998).


\bibitem{upsilon} C. Edwards et al., 
{\em Phys. Rev. Lett.} {\bf 48}, 903 (1982); 
Crystal Ball Collaboration, {\em Phys. Lett.} B {\bf 251} 204 (1990); 
CLEO Collaboration, {\em Phys. Rev.} D {\bf 51},  2053 (1995).


\bibitem{invax} A.P. Zhitnisky, {\em Yad. Fiz.} {\bf 31}, 497 (1980)
[{\em Sov. J. Nucl. Phys.} {\bf31}, 260 (1980)];
M. Dine, W. Fischler and M. Srednicki, 
{\em Phys. Lett.} B {\bf 104}, 199 (1981).


\end{thebibliography}
\end{document}